\documentclass{aa}
\usepackage{longtable}
\usepackage{supertabular}
\usepackage{graphics}
\usepackage{epsfig}
\usepackage{rotating}
\usepackage{txfonts}
\usepackage{natbib}
\begin{document}
\title{RASS-SDSS Galaxy Cluster Survey. VII.} 
\subtitle{On the Cluster Mass to Light ratio and the Halo 
Occupation Distribution.
}
\author{P. Popesso\inst{1} \and A. Biviano\inst{2} \and H. B\"ohringer\inst{3} \and M. Romaniello\inst{1}}
\institute{ European Southern Observatory, Karl Scharzschild Strasse 2, D-85748
\and INAF - Osservatorio Astronomico di Trieste, via G. B. Tiepolo 11, I-34131, Trieste, Italy
\and Max-Planck-Institut fur extraterrestrische Physik, 85748 Garching, Germany}

\abstract
{We explore the mass-to-light ratio in galaxy clusters
and its relation to the cluster mass. We study the relations among
the optical luminosity ($L_{op}$), the cluster mass ($M_{200}$) and
the number of cluster galaxies within $r_{200}$ ($N_{gal}$) in a
sample of 217 galaxy clusters with confirmed 3D overdensity. We
correct for projection effects, by determining the galaxy surface
number density profile in our cluster sample. This is best fitted by a
cored King profile in low and intermediate mass systems. The core
radius decreases with cluster mass, and, for the highest mass
clusters, the profile is better represented by a generalized King
profile or a cuspy Navarro, Frenk \& White profile. We find a very
tight proportionality between $L_{op}$ and $N_{gal}$, which, in turn,
links the cluster mass-to-light ratio to the Halo Occupation
Distribution $N_{gal}$ vs. $M_{200}$. After correcting for projection
effects, the slope of the $L_{op}-M_{200}$ and $N_{gal}-M_{200}$
relations is found to be $0.92\pm0.03$, close, but still significantly
less than unity. We show that the non-linearity of these relations
cannot be explained by variations of the galaxy luminosity
distributions and of the galaxy M/L with the cluster mass. We
suggest that the nonlinear relation between number of galaxies and
cluster mass reflects an underlying nonlinear relation between
number of subhaloes and halo mass.}
\authorrunning{P. Popesso et al.}
\maketitle

\section{Introduction}
Clusters of galaxies are the most massive gravitationally bound
systems in the universe. The cluster mass function and its evolution
provide constraints on the evolution of large-scale structure and
important cosmological parameters such as $\Omega_m$ and $\sigma
_8$. Cluster mass-to-light ratios ($M/L$ hereafter) provide one of the
most robust determination of $\Omega_m$ in connection with the
observed luminosity density in the Universe via the Oort (1958)
method. In this method, a fundamental assumption is that the average
$M/L$ of clusters is a fair representation of the universal value.
For this reason, many works have focussed on the dependence of the
cluster $M/L$ on the mass of the systems.  In general, $M/L$ has been
found to increase with the cluster mass. Assuming a power-law relation
$M/L \propto M^{\alpha}$, and adopting the usual scaling relations
between mass and X-ray temperature or velocity dispersion, when
needed, most authors have found $\alpha$ in the range 0.2-0.4, in both
optical and near-infrared bands, and over a large mass range (Adami et
al. 1998a; Bahcall \& Comerford 2002; Girardi et al. 2002; Lin et
al. 2003, 2004; Rines et al. 2004; Ramella et al. 2004; see however
Kochanek et al. 2003 for a discordant result). Why does the cluster
$M/L$ increase with the mass? Based on the results of numerical
simulations, Bahcall \& Comerford (2002) have proposed that the trend
of $M/L$ with mass is caused by the stellar populations of galaxies in
more massive systems being older than the stellar populations of
galaxies in less massive systems.  In this scenario, the slope of the
$M/L-M$ relation should be steeper in the $B$ and $V$ bands, dominated
by the young stellar populations, than at longer wavelengths,
eventually becoming flat in the infrared $K$ band, dominated by the
light of the old stellar population.  Such a scenario is not
consistent with the results of the semi-analytical modeling of
Kauffmann et al.  (1999), where the $M/L$ is predicted to increase
with mass with approximately the same slope in the $B$ and $I$
band. Also observationally, the slope of the $M/L-M$ relation is found
to be the same in different bands, the $B$-band (Girardi et al. 2002)
the $V$-band (Bahcall \& Comerford 2002), the $R$-band (Adami et
al. 1998a, Popesso et al. 2005b,2005c) and the $K$-band (Lin et
al. 2003, 2004; Rines et al. 2004; Ramella et al. 2004).  An
alternative interpretation of the increasing $M/L$ with system mass is
provided by Springel \& Hernquist (2003). They analyze the star
formation efficiency within halos extracted from cosmological
simulations, with masses in the range $10^8-10^{15} M_{\odot}$, and
find that the integrated star formation efficiency decreases with
increasing halo mass by a factor 5--10 over the cluster mass
range. This scenario is investigated by Lin et al. (2003), who convert
the 2MASS $K$-band cluster luminosities into cluster stellar
masses. They find that the fraction of mass in stars is a decreasing
function of the cluster mass ($M_{star}/M_{tot} \propto
M_{tot}^{-0.26}$).

In this paper we address the above issues by studying $M/L$ for a
sample of 217 clusters, which span the entire cluster mass range. In
particular, we study the relations among the cluster optical
luminosity $L_{op}$, the mass $M_{200}$, and the number of cluster
galaxies $N_{gal}$, within the virial radius $r_{200}$. We find a very
tight relation between $L_{op}$ and $N_{gal}$, which links the
$L_{op}-M_{200}$ relation (and therefore, the cluster $M/L$), to the
Halo Occupation Distribution (HOD hereafter) $N_{gal}-M_{200}$.  The
HOD is a powerful tool for describing galaxy bias and modelling galaxy
clustering (e.g. Ma \& Fry 2000; Peacock \& smith 2000; Seljak 2000;
Scoccimarro et al. 2001; Berlind \& Weinberg 2002). It characterizes
the bias between galaxies and mass in terms of the probability
distribution $\rm{P(N|M)}$ that a halo of virial mass M contains N
galaxies of a given type, together with relative spatial and velocity
distributions of galaxies and dark matter within halos. The HOD is a
fundamental prediction of galaxy formation theory (e.g. Kauffmann,
Nusser \& Steinmetz 1997, Kauffmann et al. 1999; White, Hernquist \&
Springel 2001; Yoshikawa et al. 2001; Berlind et al. 2003; Kravtsov et
al. 2004; Zheng et al. 2005) and it can be extremely useful to compare
the observational results with the theoretical models.

This paper is organized as follows. In section 2 we describe our
dataset. In section 3 we describe the methods we use to calculate
several cluster properties, like the characteristic radius, the virial
mass, the optical luminosity, and the number density profile of
cluster galaxies. In section 4 we analyze the $L_{op}-M_{200}$ and the
$N_{gal}-M_{200}$ relations, and find that the number of galaxies per
given halo mass decreases as the halo mass increases.  In section 5 we
seek for a physical explanation of this trend, also by comparing our
results with theoretical predictions. Finally, in section 6 we provide
our conclusions.

Throughout this paper, we use $H_0=70$ km s$^{-1}$ Mpc$^{-1}$ in a
flat cosmology with $\Omega_0=0.3$ and $\Omega_{\Lambda}=0.7$
(e.g. Tegmark et al. 2004).

\section{The data}
\label{s-data}
The optical data used in this paper are taken from the Sloan Digital
Sky Survey (SDSS, Fukugita 1996, Gunn et al. 1998, Lupton et al. 1999,
York et al. 2000, Hogg et al. 2001, Eisenstein et al. 2001, Smith et
al. 2002, Strauss et al. 2002, Stoughton et al.  2002, Blanton et
al. 2003 and Abazajian et al.  2003).  The SDSS consists of an imaging
survey of $\pi$ steradians of the northern sky in the five passbands
$u, g, r ,i, z,$ in the entire optical range.  The imaging survey is
taken in drift-scan mode.  The imaging data are processed with a
photometric pipeline (PHOTO, Lupton et al. 2001) specially written for
the SDSS data.  For each cluster we defined a photometric galaxy
catalog as described in Section 3 of Popesso et al. (2004, see also
Yasuda et al. 2001).  For the analysis in this paper we use only SDSS
Model magnitudes. The discussion about completeness limits in
magnitude and surface brightness of the SDSS galaxy photometric sample
can be found in Popesso et al. (2005a, 2005b, papers II and IV of this
series).

The spectroscopic component of the survey is carried out using two
fiber-fed double spectrographs, covering the wavelength range
3800--9200 \AA, over 4098 pixels. They have a resolution
$\Delta\lambda/\lambda$ varying between 1850 and 2200, and together
they are fed by 640 fibers, each with an entrance diameter of 3
arcsec. The fibers are manually plugged into plates inserted into the
focal plane; the mapping of fibers to plates is carried out by a
tiling algorithm (Blanton et al. 2003) that optimizes observing
efficiency in the presence of large-scale structure.

\begin{figure}
\begin{center}
\begin{minipage}{0.48\textwidth}
\resizebox{\hsize}{!}{\includegraphics{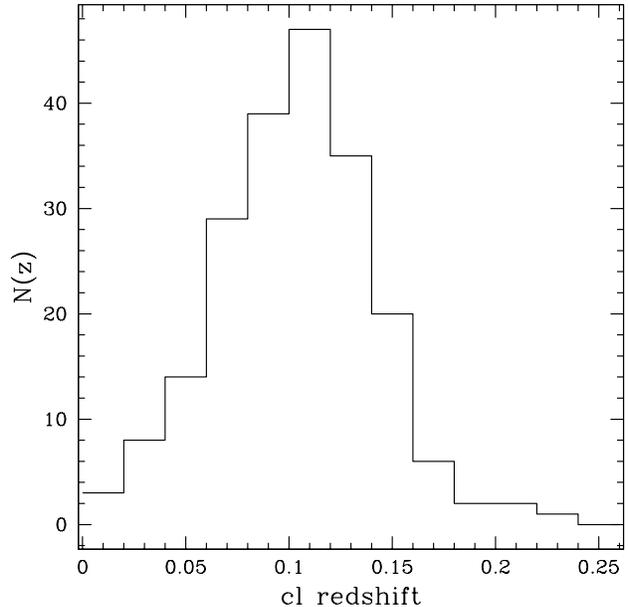}}
\end{minipage}
\end{center}
\caption{ Redshift distribution of the cluster sample used in this paper.}
\label{z_dist}
\end{figure}

\subsection{The cluster sample}
In this paper we use a combined sample of X-ray selected galaxy
clusters and optically selected systems.  The X-ray selected clusters
are taken from the RASS-SDSS galaxy cluster catalog of Popesso et
al. (2005b, hereafter paper III), which comprises 130 systems selected
mainly from the REFLEX and NORAS X-ray cluster catalogs. The optically
selected clusters are taken from Popesso et al. (2005c), who selected
a subsample of 130 Abell clusters with confirmed 3d galaxy overdensity
in the third release of the SDSS galaxy spectroscopic catalog. The two
samples overlap with 43 clusters. The combined sample with the
exclusion of the double detections comprises 217 clusters and covers
the entire range of masses and X-ray/optical luminosities, from very
low-mass and X-ray/optical faint groups ($10^{13} M{\odot}$) to very
massive and X-ray/optical bright clusters ($5\times10^{15}
M{\odot}$). The cluster sample comprises only nearby systems at the
mean redshift of 0.1. The redshift distribution of the cluster
sample is shown in Fig. \ref{z_dist}.

\section{The cluster properties}
In this section we explain the methods used to calculate the cluster
properties as the characteristic radius, the virial mass, the optical
luminosity and the parameters of the radial profile of the cluster
galaxies.

\subsection{Characteristic radii and masses}
We here describe the methods by which we measure the characteristic
cluster radii $r_{200}$ and mass $M_{200}$. $r_{200}$ and $M_{200}$ are
the radius and the mass, respectively, where the mass density of the
system is 200 times the critical density of the Universe and it is
considered as a robust measure of the virial radius of the cluster.

Estimates of a cluster velocity dispersion, mass, and characteristic
radius requires knowledge of the redshifts of its member galaxies. We have 
used the redshifts provided in the SDSS spectroscopic catalog.

Cluster members are selected following the method of Adami et
al. (1998a) or Girardi et al. (1993), depending on whether the mean
cluster redshift $z_{cluster}$ is known in advance (from previous
studies) or not, respectively. Girardi et al. (1993) method requires
in fact that a preliminary cut be done in the line-of-sight velocity
space, $\pm 4000$~km~s$^{-1}$ around $c z_{cluster}$, before searching
for significant weighted-gaps in the velocity distribution. On the
other hand, the density-gap technique of Adami et al. does not require
such a preliminary cut to be operated. If $z_{cluster}$ is known
already, we select among the groups identified by the gapping
technique that one closest in velocity space to $c z_{cluster}$,
otherwise we select the most populated one. After the initial group
selection, we apply the interloper-removal method of Katgert et
al. (2004; see Appendix A in that paper for more details) on the
remaining galaxies, using the X-ray center when available, or else
the position of the brightest cluster galaxy on the cluster
colour-magnitude sequence.

The virial analysis (see, e.g., Girardi et al. 1998) is then performed
on the clusters with at least 10 member galaxies. The velocity dispersion
is computed on the cluster members, using the biweight estimator
(Beers et al. 1990). The virial masses are corrected for the surface
pressure term (The \& White 1986) by adopting a profile of Navarro et
al.  (1996, 1997; NFW hereafter) with a concentration parameter, $c$,
that depends on the initial estimate of the cluster virial mass
itself.  The $c$--mass relation is given by $c=4 \times
(M/M_{KBM})^{-0.102}$ where the slope of the relation is taken from
Dolag et al. (2004), and the normalization $M_{KBM} \simeq 2 \times
10^{15} M_{\odot}$ from Katgert et al. (2004). The clusters in our
sample span a range $c \simeq 3$--6.

Correction for the surface pressure term requires knowledge of the
$r_{200}$ radius, for which we adopt Carlberg et al. (1997)
definition (see eq.(8) in that paper) as a first guess. After the
virial mass is corrected for the surface pressure term, we refine our
$r_{200}$ estimate using the virial mass density itself. Say $M_{vir}$
the virial mass (corrected for the surface term) contained in a volume
of radius equal to a chosen observational aperture, $r_{ap}$. The
radius $r_{200}$ is then given by:
\begin{equation}
r_{200} \equiv r_{ap} \, [\rho_{vir}/(200 \rho_c)]^{1/2.4}
\label{e-r200}
\end{equation}
where $\rho_{vir} \equiv 3 M_{vir}/(4 \pi r_{ap}^3)$ and $\rho_c(z)$
is the critical density at redshift $z$ in the adopted cosmology. The
exponent in eq.(\ref{e-r200}) is the one that describes the average
cluster mass density profile near $r_{200}$, as estimated by Katgert
et al. (2004) for an ensemble of 59 rich clusters. 

A NFW profile is used to interpolate (or, in a few cases, extrapolate)
the virial mass $M_{vir}$ from $r_{ap}$ to $r_{200}$, yielding
$M_{200}$. As before, we scale the concentration parameter of the used
NFW profile according to a preliminary estimate of the mass of the
system. From $M_{200}$ the final estimate of $r_{200}$ is obtained,
using the definition of $M_{200}$ itself.

\subsection{Optical luminosities}
The total optical luminosity of a cluster has to be computed after the
subtraction of the foreground and background galaxy contamination. We
consider two different approaches to the statistical subtraction of
the galaxy background. We compute the local background number counts
in an annulus around the cluster and a global background number counts
from the mean of the magnitude number counts determined in five
different SDSS sky regions, randomly chosen, each with an area of 30
$\rm{deg^2}$. In our analysis we show the results obtained using the
optical luminosity estimated with the second method. The optical
luminosity is then computed within $r_{200}$ following the
prescription of Popesso et al. (2004). The reader is referred to that
paper for a detailed discussion about the comparison between optical
luminosities calculated with different methods.  To avoid selection
effects due to the slightly different redshifts of the clusters, the
optical luminosity has been calculated in the same absolute magnitude
range for all the clusters. The adopted range has been varied to check
the robustness of the results of the regression analyses.

\begin{figure}
\begin{center}
\begin{minipage}{0.49\textwidth}
\resizebox{\hsize}{!}{\includegraphics{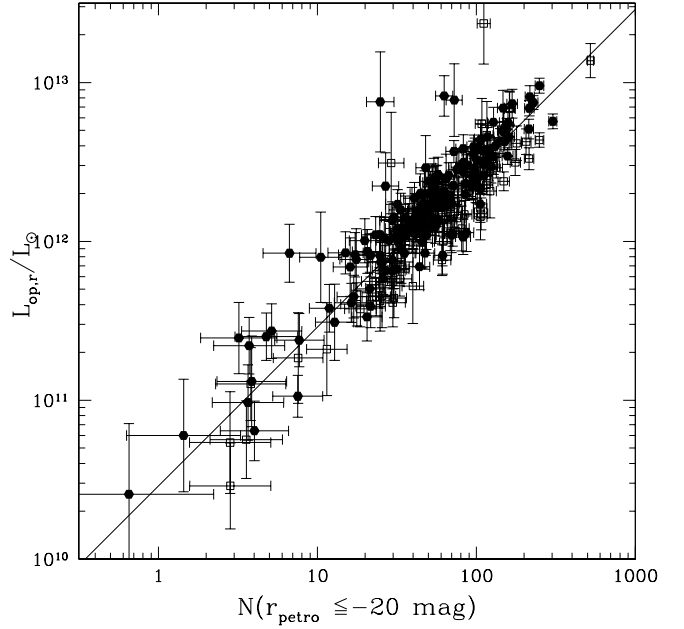}}
\end{minipage}
\end{center}
\caption{
Relation between the optical luminosity calculated in the SDSS r band
within $r_{200}$ and the number of cluster galaxies contributing to
$L_{op}$. The solid line is the best fit line with slope
$1.00\pm0.03$.}
\label{z_gal}
\end{figure}

\subsection{Number density profiles and projection effects}\label{dprofs}
The observed cluster optical luminosity, $L_{op}$, is contributed not
only by galaxies within the virial sphere of radius $r_{200}$, but
also by galaxies outside the virial sphere yet within the cylinder of
same radius. It is therefore necessary to correct the observed optical
luminosity for the contribution of cluster galaxies outside the virial
sphere (the field galaxies contribution is removed statistically as
described in the previous section).  

Fig. \ref{z_gal} shows the proportionality between the cluster r-band
optical luminosity within $r_{200}$ and the number of cluster galaxies
($N_{gal}$), contributing to the luminosity itself, i.e. the
background-subtracted galaxy counts within the same radius, down to
the magnitude limit used to estimate $L_{op}$. Because of the strict
proportionality between $L_{op}$ and $N_{gal}$, we can use the ratio
between the number of cluster galaxies within the cylindrical volume
and the number of galaxies within the virial sphere, to correct the
observed $L_{op}$ for the contribution of cluster galaxies outside the
virial sphere. In order to estimate this ratio, we build the surface
number density profiles of our clusters, and fit them with two
widely-used analytical functions, the King (1962) cored profile, and
the NFW cuspy profile. The 3D and projected King profiles are given
by, respectively:
\begin{equation}
\rho(r)=\frac{n_0}{(1+(r/r_c)^2)^{3/2}}
\label{king1}
\end{equation}
and:
\begin{equation}
\sigma(b)=\frac{\sigma_0}{(1+(b/r_c)^2)}
\label{king2}
\end{equation}
where $r_c$ is the core radius and $\sigma_0=2n_0r_c$ is the
normalization (see also Sarazin 1980).  The NFW profile in 3D is given by:
\begin{equation}
\rho(r)=\frac{\delta_0}{r/r_s(1+(r/r_s)^2)}
\end{equation}
where $r_s$ is the characteristic radius ($r_s=r_{200}/c$ with $c$
the concentration parameter) and $\delta_0$ is the normalization. The
projected surface density profile is then obtained from an integration
of the three-dimensional profile (see Bartelmann et al. 1996).

\begin{figure}
\begin{center}
\begin{minipage}{0.49\textwidth}
\resizebox{\hsize}{!}{\includegraphics{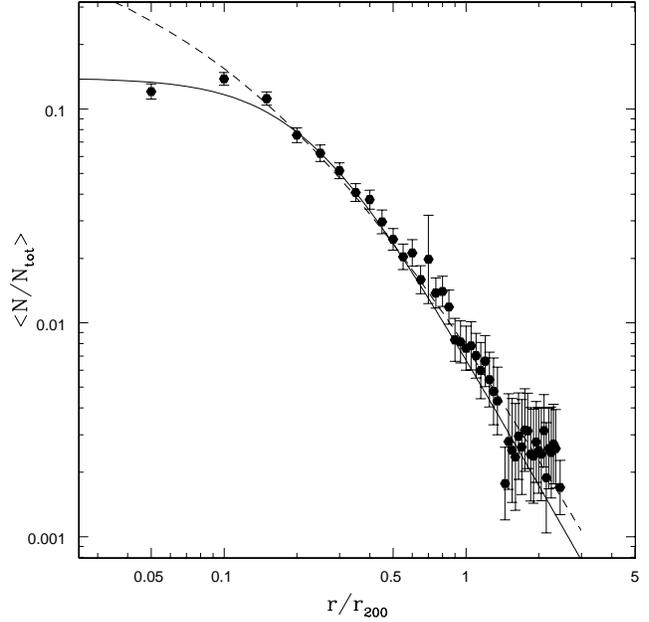}}
\end{minipage}
\end{center}
\caption{ The stacked mean surface number density profile of all the
cluster galaxies with magnitude $r < -18.5$.  The solid curve is
the best fit King profile, the dashed curve is the best fit NFW
profile.}
\label{mean}
\end{figure}

As a first step we explore the the mean surface density galaxy
distribution within our cluster sample, by stacking the projected
galaxy distributions of the individual systems. Note that in this
analysis we only consider the clusters with available X-ray centers,
in order to reduce possible mis-centering when adopting the positions
of brightest cluster galaxies as cluster centers (not all brightest
cluster galaxies lie at centers of their parent clusters, see, e.g.,
Lin \& Mohr 2004).  The clustercentric distances are rescaled to the
cluster $r_{200}$ before the stacking.  The cluster galaxy
distributions are normalized to the total number of galaxies within
$r_{200}$, after subtraction of the mean background galaxy density,
evaluated within the $2.5-3.5\times r_{200}$ annulus. Fig. \ref{mean}
shows the stacked surface density profile of all the 217 clusters. The
best fit is given by a King profile with core radius
$r_c/r_{200}=0.224\pm0.005$, while a NFW profile provides a poor fit
near the centre.  We then split our sample of clusters in 6 mass bins:
$M_{200}/10^{14} M_{\odot} \le 1$, $1 < M_{200}/10^{14} M_{\odot} \le
3$, $3 < M_{200}/10^{14} M_{\odot} \le 7$, $7 < M_{200}/10^{14}
M_{\odot} \le 10$,$10 < M_{200}/10^{14} M_{\odot} \le 30$, and
$M_{200}/10^{14} M_{\odot} > 30$.  Each bin contains at least 10
clusters.

\begin{figure*}
\begin{center}
\begin{minipage}{0.8\textwidth}
\resizebox{\hsize}{!}{\includegraphics{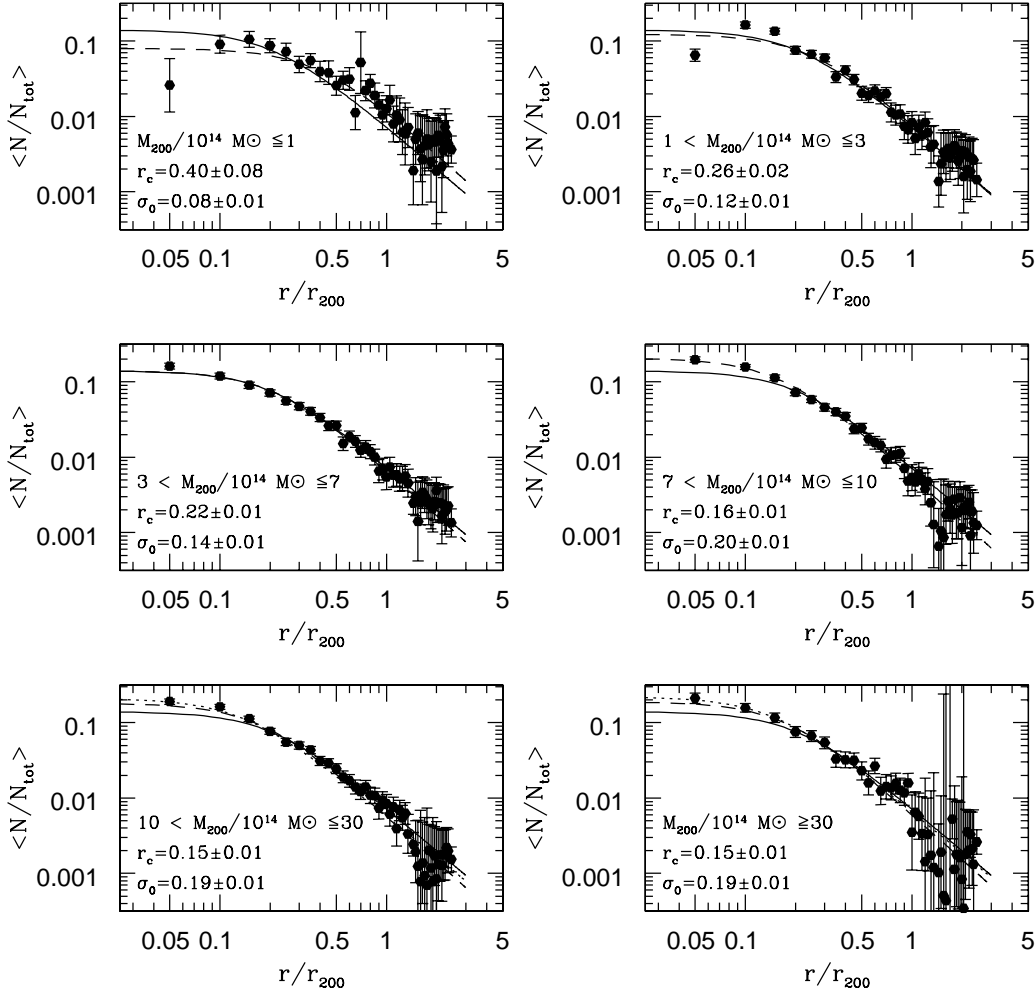}}
\end{minipage}
\end{center}
\caption{
The stacked surface number density profiles of clusters in different
cluster mass bins. The individual cluster profiles are obtained by
considering all the galaxies with $r_{petro} < -18.5$.  The dashed
curve in each panel is the King profile that provides the best-fit to
the surface density profile shown in that same panel, while the solid
curve is the King profile that provides the best-fit to the mean
stacked surface density profile (same as the solid line in
Fig.~\ref{mean}).  The dotted line in the bottom panels (corresponding
to the highest-mass bin clusters) is the best fit provided by a
generalized King profile.}
\label{plotto}
\end{figure*}

\begin{figure}
\begin{center}
\begin{minipage}{0.49\textwidth}
\resizebox{\hsize}{!}{\includegraphics{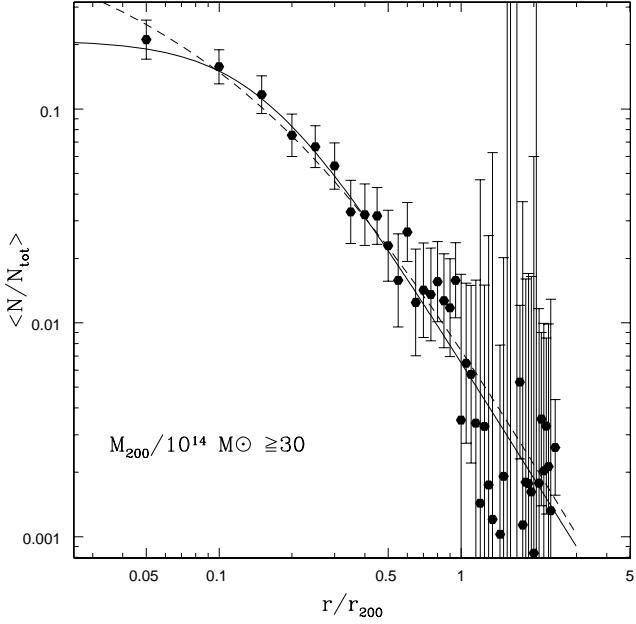}}
\end{minipage}
\end{center}
\caption{
The surface number density profile of all the cluster galaxies with
$r < -18.5$ in the highest of our considered cluster mass
bins. The solid curve is the best fit provided by the generalized King
profile. The dashed curve is the best fit provided by the NFW profile.
Both profiles are consistent with the data.
}
\label{prof7}
\end{figure}

Fig. \ref{plotto} shows the surface density profiles in each cluster
mass bin. The solid line in each plot shows the King profile that
provides the best-fit to the surface density profile of all galaxies
in all clusters, already shown in Fig.~\ref{mean}. The dashed line in
each panel is the best-fit King profile for the surface density
profile of each cluster mass bin. The NFW profiles provide poor fits
for most cluster mass bins, and are not plotted.  From
Fig.~\ref{plotto} one can clearly see how the cluster galaxy
distribution changes with cluster mass. The surface density profiles
become steeper near the centre as the cluster mass increases. Note in
fact that the surface number density profile in the low mass bin
($M_{200}/10^{14} M_{\odot} \le 1$) is not completely consistent with
a King profile since it shows a deficit of galaxies near the
center. The core radius is quite large, $r_c/r_{200}=0.40\pm0.08$. The
core radius becomes smaller as the cluster mass increases, and it is
$r_c/r_{200}=0.16\pm0.01$ for clusters in the mass interval $7 <
M_{200}/10^{14} M_{\odot} \le 10$. In the last two mass bins, the
galaxy distributions become so concentrated that the simple King
profile does no longer provide a good fit, and a generalized King
profile is needed, of the form:
\begin{equation}
\sigma(b)=\frac{\sigma_0}{(1+(b/r_c)^2)^{\beta}}.
\end{equation}
The dotted lines in the panels of Fig. \ref{plotto} corresponding to
the highest cluster mass bins, show the best fit given by the
generalized King profile, where $\beta=0.91\pm0.01$. Finally, in the
highest mass bin ($M_{200} > 3 \times 10^{15}$) also the projected NFW
profile provides a good fit to the galaxy distributions. This is shown
in Fig.~\ref{prof7}, where the best-fit generalized King profile is
shown as a solid curve, and the best-fit NFW profile is shown as a
dashed curve. In this case, the best-fit value of the NFW
concentration parameter is $c=4.2\pm0.3$, and is consistent with the
value found for the dark matter distribution in similarly massive
clusters (e.g. Biviano \& Girardi 2003; Katgert et al. 2004).

Lin et al. (2004) perform the same analysis on a smaller sample of 93
X-ray selected clusters observed in the 2MASS all sky survey. Their
conclusion is that the surface density profile of cluster galaxies is
consistent with a NFW profile with concentration parameter
$2.90\pm0.22$. They study the galaxy distribution in 2 mass bins with
mean mass $<M_{500}>=7.9\times 10^{13}M{\odot}$ for the groups, and
$<M_{500}>=5.3\times 10^{14}M{\odot}$, for the massive clusters, and
claim that the spatial profiles are consistent with the mean profile
in both mass bins. However, by fitting their data (taken from Fig.~8
of Lin et al. 2004) with both a King, a generalized King, and a NFW
profile, we find that a King profile provides the best fit, in
agreement with our findings.

\begin{figure}
\begin{center}
\begin{minipage}{0.49\textwidth}
\resizebox{\hsize}{!}{\includegraphics{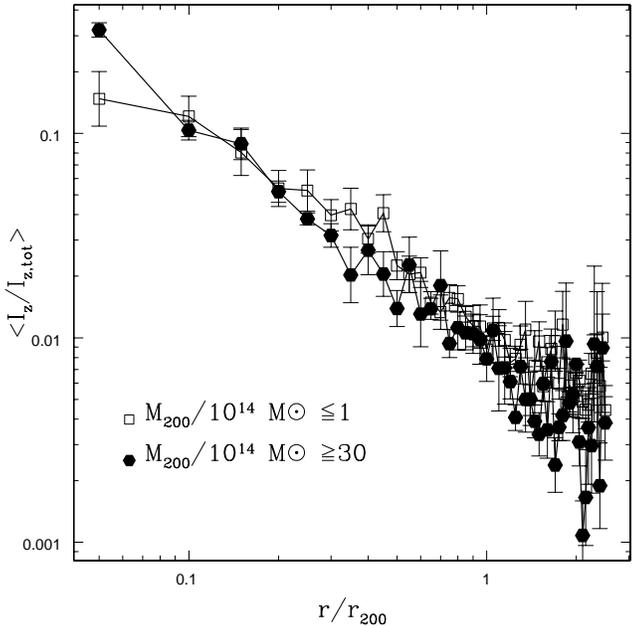}}
\end{minipage}
\end{center}
\caption{
Composite surface brightness profile of two cluster samples: the
low mass systems at $M_{200} \le 10^{14} M_{\odot}$ (empty squares)
and the massive clusters at $M_{200} > 3\times 10^{15} M_{\odot}$
(filled dots).}
\label{lum_profile.eps}
\end{figure}

Our results are further supported by the analysis of the surface
brightness profile of our clusters. Fig. \ref{lum_profile.eps} shows
the composite surface brightness profile of two cluster subsamples:
the low-mass systems with $M_{200} \le 10^{14} M_{\odot}$ and the
high-mass clusters with $M_{200} > 3\times 10^{15} M_{\odot}$. The
profile of the low-mass objects displays a core, and is less centrally
concentrated than that of the high-mass clusters which is in fact
rather cuspy. As expected, due to the presence of the Brightest
Cluster Galaxies at the center of the systems, the luminosity profiles
are generally cuspier than the density profiles in the same cluster
mass bins (Adami et al. 2001).

In stacking clusters, we have assumed circularity, because the
number of galaxies per cluster is generally too small to allow a
precise determination of individual cluster shapes and orientations.
Adami et al. (1998b) have shown that enforcing circularity could
create a central artificial cusp in the number density profile of the
stacked cluster. However, lower mass clusters are more elongated than
higher mass clusters (see Fasano et al. 1993; de Theije et al. 1995;
Plionis et al. 2004), so the effect of assuming circularity should
lead to cuspier density profiles for lower mass clusters, which is
opposite to what we find. Indeed, the effect reported by Adami et
al. does not seem to be strong enough to account for the differences
seen in the density profiles of the stacked clusters of different
masses (compare Fig.~\ref{plotto} with Fig.~7 in Adami et al. 1998b).

Hence we conclude that there is a significant variation of the number
density and luminosity density profiles of clusters, as a function of
cluster mass, with higher mass clusters displaying more concentrated
profiles.  As a consequence, also the correction needed to convert the
number of galaxies contained in the cylindrical volume to that in the
virial sphere depends on the cluster mass.  Using the volume and the
surface density King profile given in eq. \ref{king1} and \ref{king2},
respectively, we estimate that the ratio between the number of
galaxies in the virial sphere of radius $r_{200}$ and the number of
galaxies actually observed in the cylinder of same radius is
0.69--0.76 for clusters in the lowest-mass bin, 0.78--0.80 for
clusters in the $1-3 \times 10^{14}M_{\odot}$ mass bin, 0.81 in the
$3-7 \times 10^{14}M_{\odot}$ mass bin, and 0.85 in the highest mass
bins.

\begin{figure*}
\begin{center}
\begin{minipage}{0.8\textwidth}
\resizebox{\hsize}{!}{\includegraphics{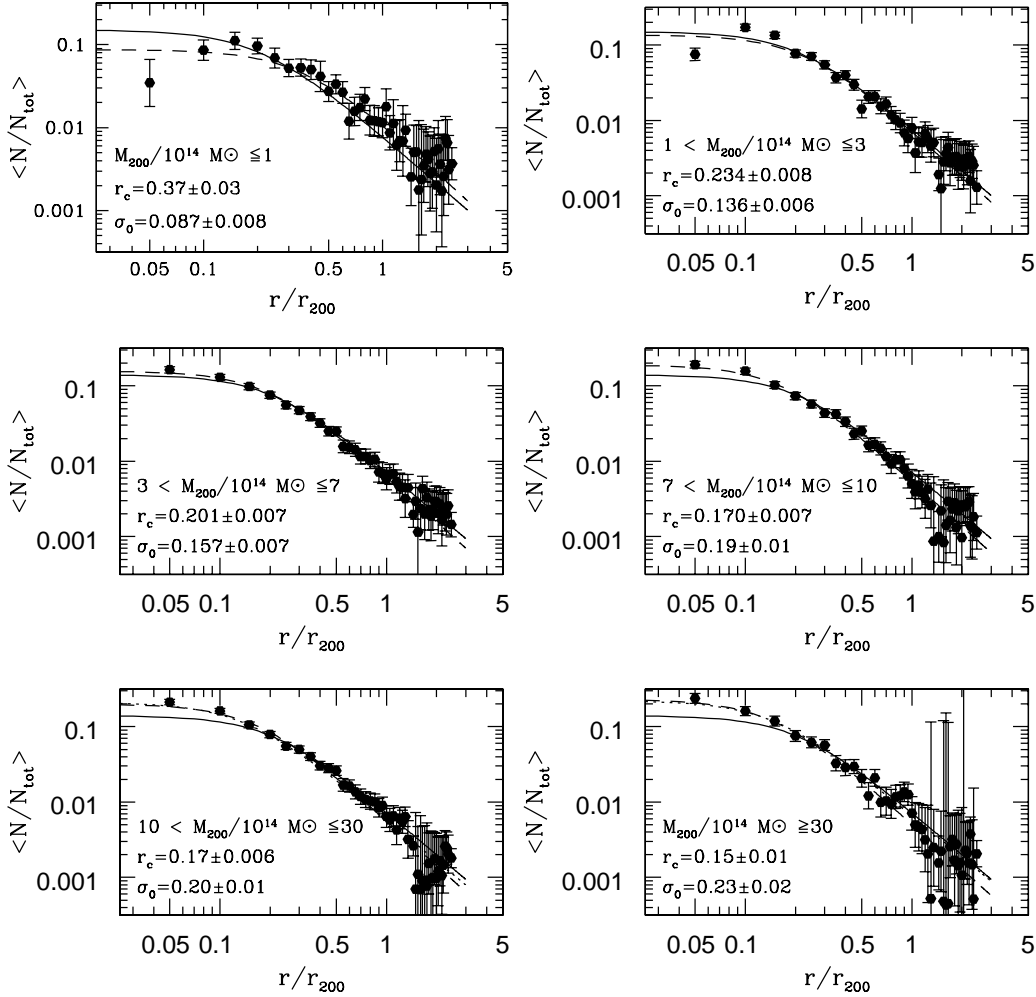}}
\end{minipage}
\end{center}
\caption{
The stacked surface number density profile of 
the {\em red} cluster galaxies with
magnitude $r < -18.5$, separately for clusters of different masses.
The meaning of the lines is the same as in Fig.~\ref{plotto}.}
\label{plottored}
\end{figure*}

\begin{figure*}
\begin{center}
\begin{minipage}{0.8\textwidth}
\resizebox{\hsize}{!}{\includegraphics{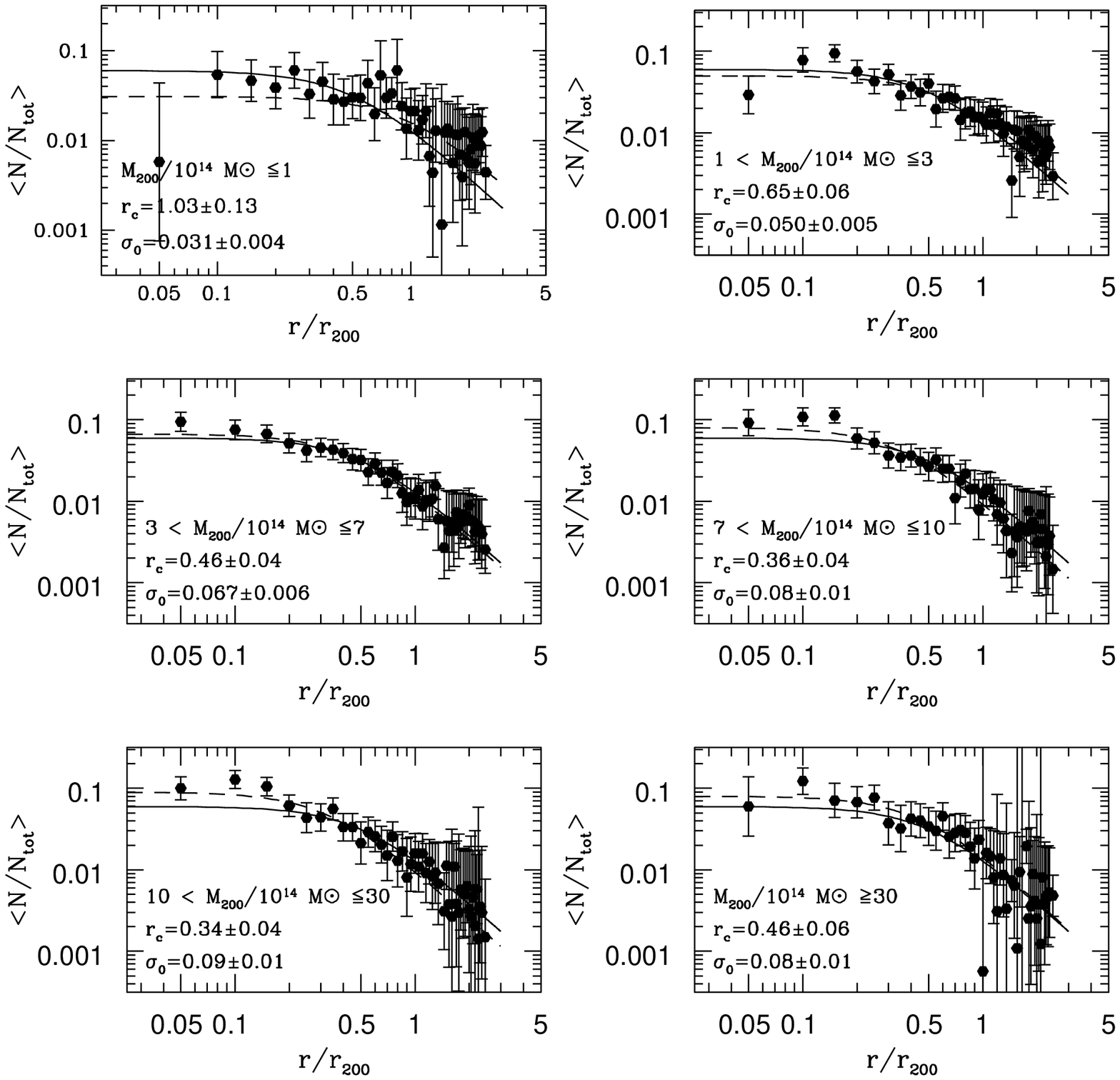}}
\end{minipage}
\end{center}
\caption{
The stacked surface number density profile of 
the {\em blue} cluster galaxies with
magnitude $r < -18.5$, separately for clusters of different masses.
The meaning of the lines is the same as in Fig.~\ref{plotto}.}
\label{plottoblue}
\end{figure*}

We performed the same analysis separately for the red (early-type) and
blue (late-type) cluster galaxy populations. The colour separation
between the two population is based on the SDSS galaxy color u-r
(Strateva et al. 2001, Popesso et al 2006).  For both the red and the
blue galaxy populations, the core radius of the best-fit King profile
monotonically decreases from the low-mass systems to the more massive
clusters (see Figure~\ref{plottored} and \ref{plottoblue}).

Since there is a significant mass-dependence of the number density
profiles, a mass-dependent deprojection correction needs to be
applied to the observed values of $L_{op}$. In the following, we
only consider the deprojection-corrected values of $L_{op}$, obtained
by adopting the correction factors per mass bin derived above.

\section{The $L{op}-M_{200}$ and the $N_{gal}-M_{200}$ relations}\label{lmr}
\begin{figure}
\begin{center}
\begin{minipage}{0.5\textwidth}
\resizebox{\hsize}{!}{\includegraphics{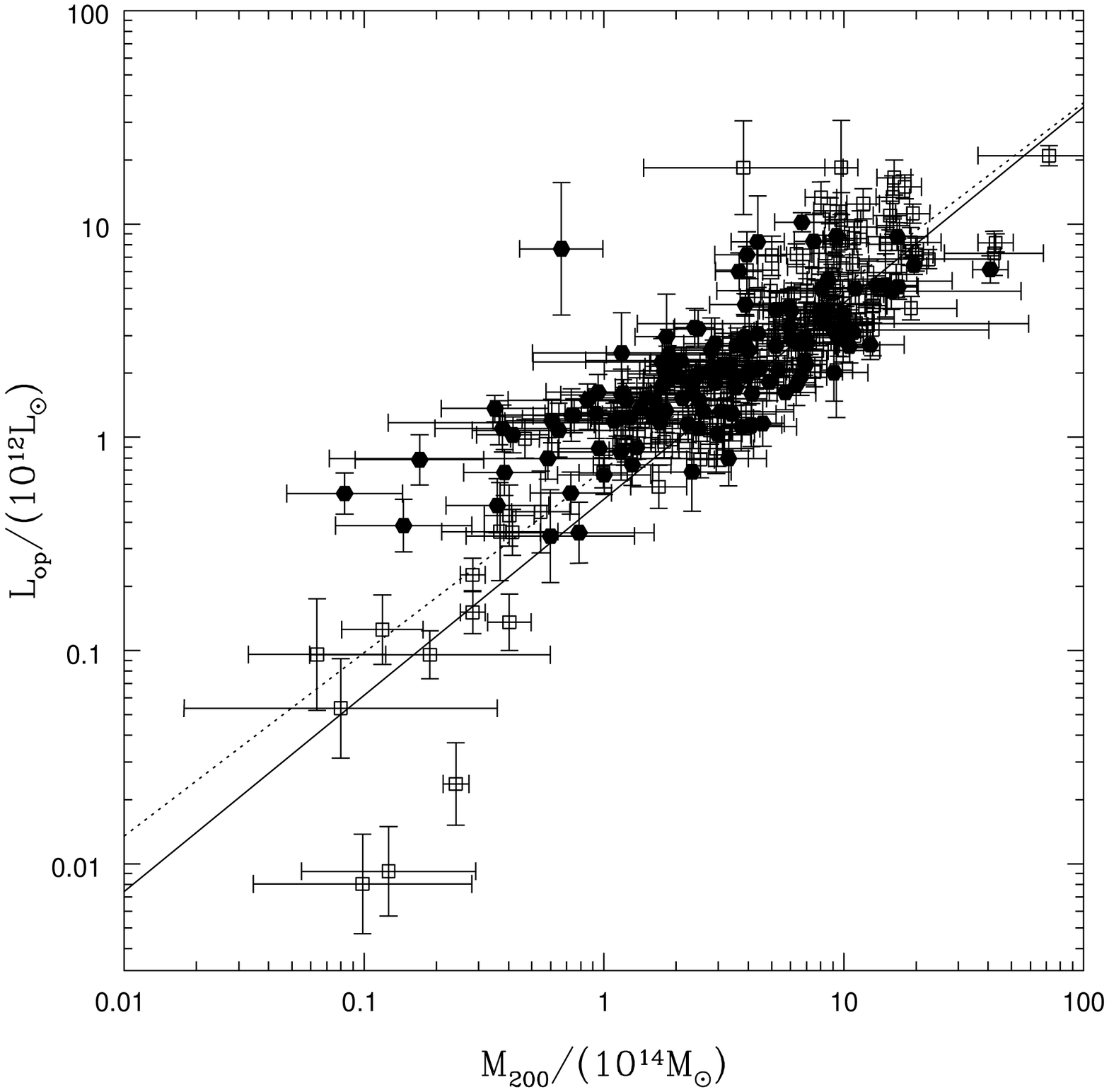}}
\end{minipage}
\end{center}
\caption{ $L_{op}-M_{200}$ relation. The optical luminosity is
calculated within $r_{200}$ and is corrected for contamination due to
projection effect. The empty squares in the plot are the X-ray
selected clusters. The filled points are the optically selected
clusters. The solid line in the plot is the best fit line of the
corrected $L_{op}-M_{200}$ relation. The dashed line is the best fit
line of the uncorrected $L_{op}-M_{200}$ relation. }
\label{lom}
\end{figure}

\begin{figure}
\begin{center}
\begin{minipage}{0.5\textwidth}
\resizebox{\hsize}{!}{\includegraphics{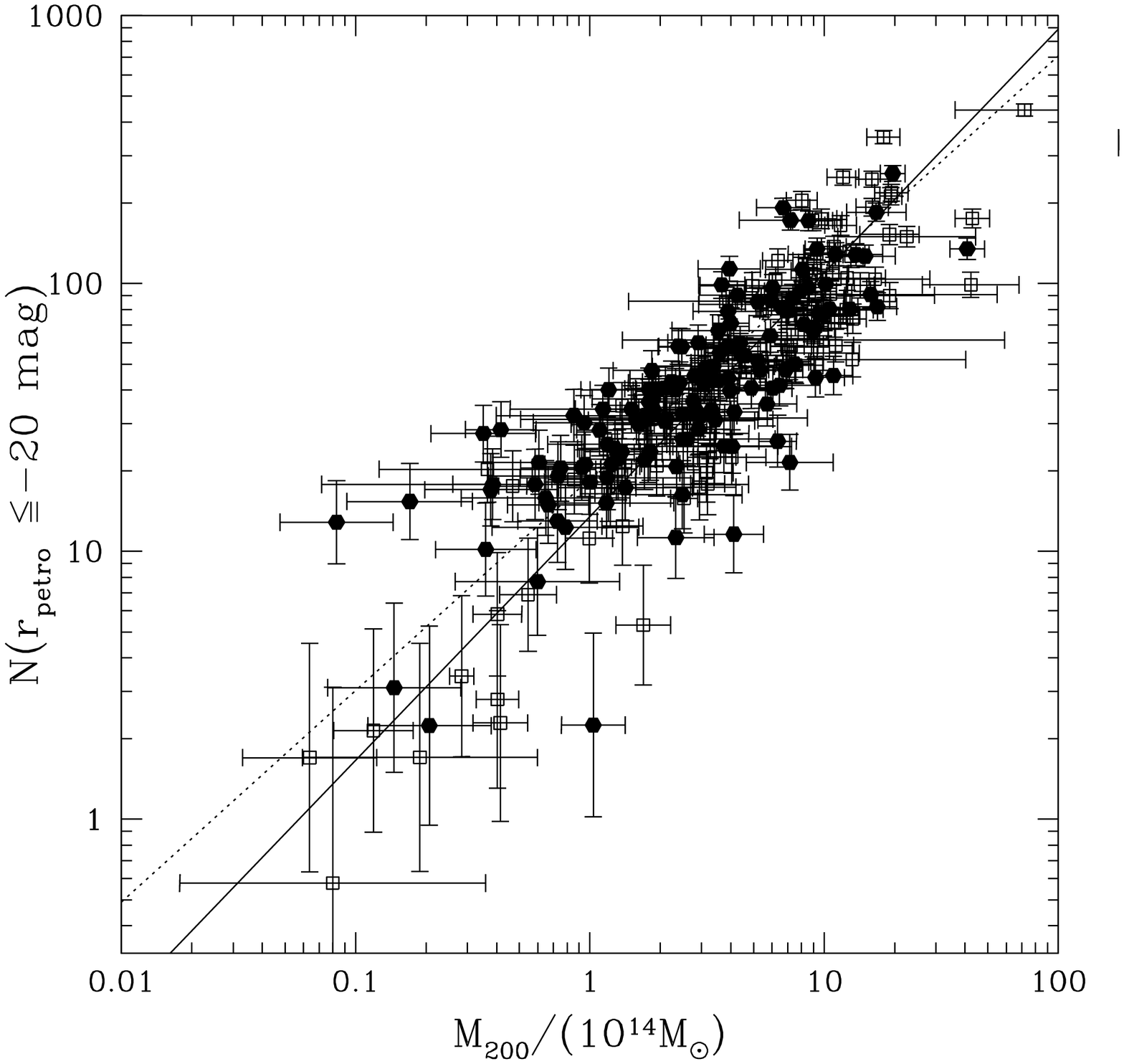}}
\end{minipage}
\end{center}
\caption{
$N_{gal}-M_{200}$ relation. The number of cluster galaxies is
calculated within $r_{200}$ and is corrected for contamination due to
projection effect. The empty squares in the plot are the X-ray
selected clusters. The filled points are the optically selected
clusters. The solid line in the plot is the best fit line of the
corrected $N_{gal}-M_{200}$ relation. The dashed line is the best fit
line of the uncorrected $N_{gal}-M_{200}$ relation.}
\label{def}
\end{figure}

In Fig. \ref{lom} we show the r-band $L_{op}-M_{200}$ relation after
correcting $L_{op}$ for the projection effects (see
section~\ref{dprofs}). The solid line in the figure is the best-fit
linear regression in logarithmic space, and the dotted line is the
best-fit we would have obtained had $L_{op}$ not been corrected for
the projection effects. The best-fit relation for the corrected
$L_{op}$ is:
\begin{equation}
L_{op}/(10^{12}L_{\odot})=10^{-0.29\pm0.03}(M_{200}/(10^{14}M_{\odot}))^{0.92\pm0.03}.
\end{equation}
The slope of this relation is steeper than the slope of the
uncorrected relation, which is $0.85\pm0.03$. The two values are
anyhow marginally consistent within $2.5\sigma$. As a consequence,
also the $M/L-M$ relation is flatter.  The slope of the corrected
relation is $0.18\pm0.04$ instead of $0.27\pm0.04$ for the uncorrected
relation. Remarkably, we find that the slopes of the best-fit $L-M$
and $M/L-M$ relations do not depend on the chosen photometric SDSS
band.

Due to the strict proportionality observed between the cluster optical
luminosity and the number of cluster galaxies (see Fig. \ref{z_gal}),
it is clear that the $L_{op}-M_{200}$ relation is strictly connected
to the $N_{gal}-M_{200}$ relation. In other words, the cluster
mass-to-light ratio $M/L$ is strictly related to the Halo Occupation
Number $\gamma$ of the Halo Occupation Distribution (HOD) $N\propto
M^{\gamma}$. It is then useful to study the cluster $M/L$ in terms of
the HOD since this allows an easier comparison with the predictions of
models of structure formation.

We study the HOD with two approaches. First we use the $N_{gal}$
calculated using the photometric data ($N_{phot}$), by summing the
background-subtracted cluster number counts used to calculate
$L_{op}$. As a second approach we estimate the number of
spectroscopically-confirmed cluster members ($N_{spec}$). Both
estimates are corrected for projection effects in the same way as we
did for $L_{op}$. Both $N_{phot}$ and $N_{spec}$ are computed down to
the same absolute magnitude, in order to allow comparison of the two
estimates. The SDSS spectroscopic and photometric catalogs have two
different apparent magnitude limits ($r= 17.77$ for the spectroscopic
catalog and $r \sim 21$ mag for the photometric one).  We apply an
absolute magnitude cut of $M_r \leq -20$, which allows us to sample
the cluster luminosity function (LF hereafter) down to $M^*+2$
(Popesso et al. 2005a). With such a cut, $N_{spec}$ can be measured
for a significant fraction of our cluster sample, those 90 clusters
for which $M_r \leq -20$ is brighter than the apparent magnitude limit
of $r= 17.77$.

In Fig. \ref{def} we show the $N_{gal}-M_{200}$
relation, using $N_{gal} \equiv N_{phot}$. We also  plot the
best fit relations
\begin{equation}
N_{gal}=10^{-11.60\pm0.59}(M_{200}/M_{\odot})^{0.91\pm0.04}
\end{equation}
for $N_{gal} \equiv N_{phot}$, and
\begin{equation}
N_{gal}=10^{-11.43\pm0.76}(M_{200}/M_{\odot})^{0.89\pm0.05}
\end{equation}
for $N_{gal} \equiv N_{spec}$.  The two estimates of the halo
occupation number $\gamma$ are consistent, while the different
normalizations reflect the incompleteness of the spectroscopic samples
(see Popesso et al. 2006).  The orthogonal scatter in both relations
is 35\%, and $M_{200}$ can be predicted from $N_{gal}$ with an
accuracy of 55\%.

Had we not corrected $N_{gal}$ for the projection effects, we would
have underestimated the slope for the $N_{gal}-M_{200}$ relation,
obtaining $0.79\pm0.04$. Clearly, applying an average, mass
independent, correction to the observed value of $N_{gal}$ and
$L_{op}$ leads to underestimate the slope of the considered relations.

We check also if different cluster selection techniques introduce
biases in our analysis. For this purpose we perform the same analysis
separately on the optically and X-ray selected cluster samples,
respectively. The observed best fit values are consistent within the
statistical errors. Moreover, we perform the analysis by adopting
different magnitude cuts to check for vatiation of the Halo Occupation
number in different magnitude regimes. We consider the following
magnitude cuts: $-20$, $-17$ and $-16$ mag in the i band. While the
normalization of the relation is obviously changing, the best fit
values of the Halo Occupation number are cosistent within the errors
in all the magniture ranges.

\begin{figure}
\begin{center}
\begin{minipage}{0.5\textwidth}
\resizebox{\hsize}{!}{\includegraphics{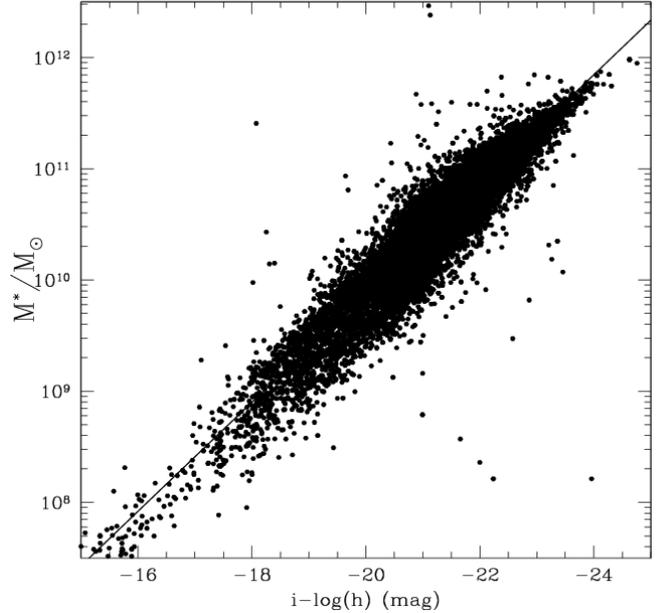}}
\end{minipage}
\end{center}
\caption{ 
Relation between the i band petrosian absolute magnitude and the galaxy
stellar mass. The galaxy stellar masses are taken by Kauffmann et
al. (2003).  }
\label{maz}
\end{figure}

The Halo Occupation Number $\gamma$ has been measured with several
different methods in the literature. Most of these come from assuming
a form of the HOD, and adjusting the parameters until the prediction
from the halo model matches the observed galaxy clustering
(e.g. Seljak et al. 2004, Peacock \& Smith 2000; Yang et al. 2003;
Zehavi et al. 2004; Magliocchetti \& Porciani 2003).  Pisani et
al.(2003) used the velocity dispersion in the groups of the Zwicky
catalog and obtained $\gamma=0.70\pm0.04$, while Marinoni \& Hudson
(2002) used the LF of the Nearby Optical Galaxy sample and obtained
$\gamma=0.55\pm0.043$.

Many other works in the literature used an approach similar to ours.
Kochanek et al. (2003) used a sample of clusters identified in the
2MASS all sky survey, and obtained $\gamma=1.11 \pm 0.09$ on a sample
of 84 clusters.  Lin et al. (2004) used a sample of 93 X-ray clusters
observed in 2MASS, and found $\gamma=0.84\pm0.04$. Similar results
were obtained by Yang et al. (2005) who used a large sample of groups
identified in the 2-degree Field Galaxy Redshift Survey.

With the exclusion of Kochanek et al. (2003), all other studies agree
on the fact that the exponent in the $N-M$ relation, and consequently
in the $L-M$ relation, is not consistent with unity (see Lin et al.
2004 for a discussion about the discrepancy with the results of
Kochanek et al. 2003). However, with the mass-dependent correction
applied to our clusters to clean the $N-M$ ($L-M$) relation from
projection effects, the estimated value of $\gamma$ becomes closer to
unity. Nevertheless, a direct proportionality between cluster mass and
number of cluster galaxies is still excluded by our analysis at the
$\sim 2$--$2.5 \sigma$ level.

\begin{figure*}
\begin{center}
\begin{minipage}{0.48\textwidth}
\resizebox{\hsize}{!}{\includegraphics{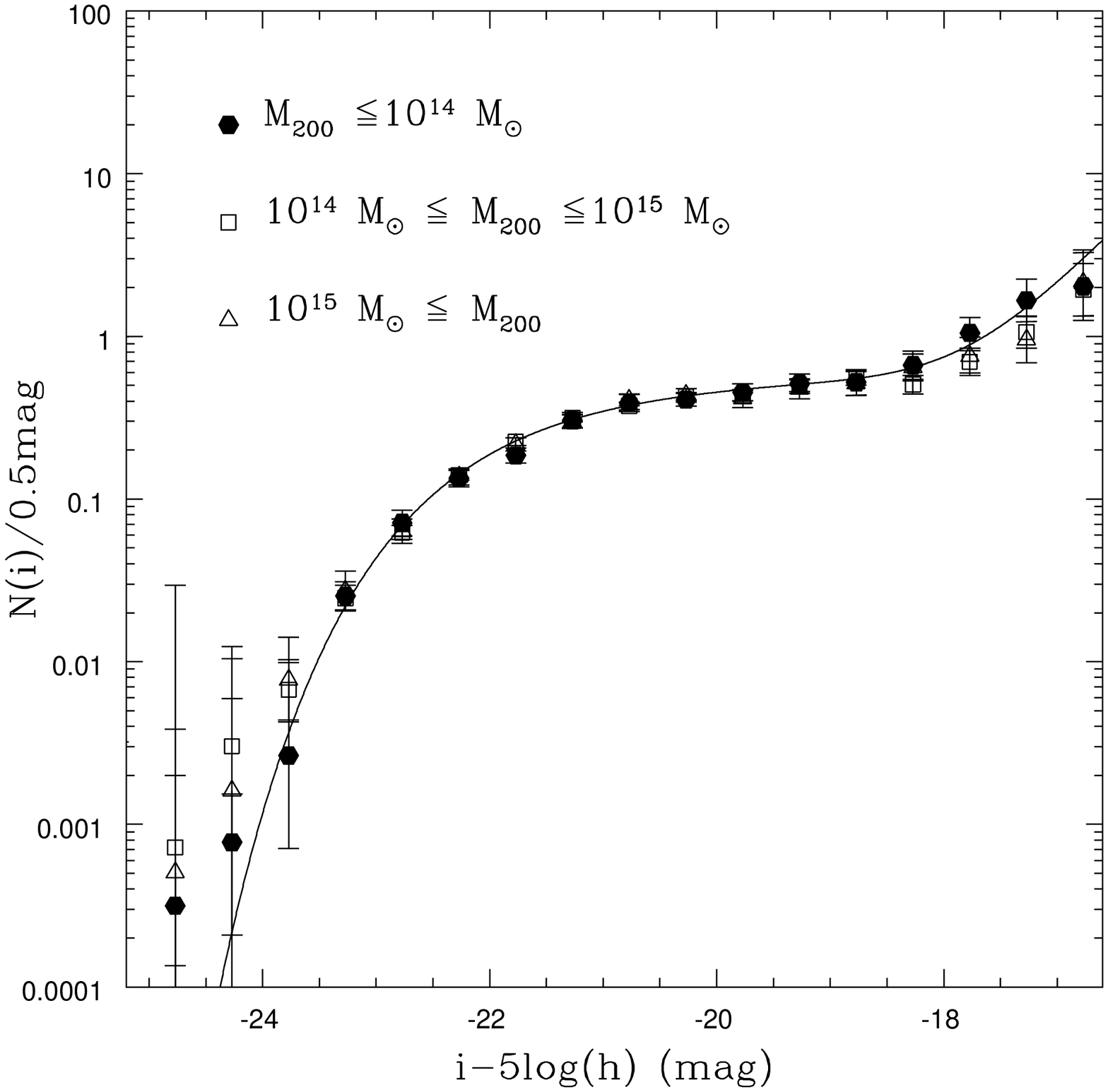}}
\end{minipage}
\begin{minipage}{0.48\textwidth}
\resizebox{\hsize}{!}{\includegraphics{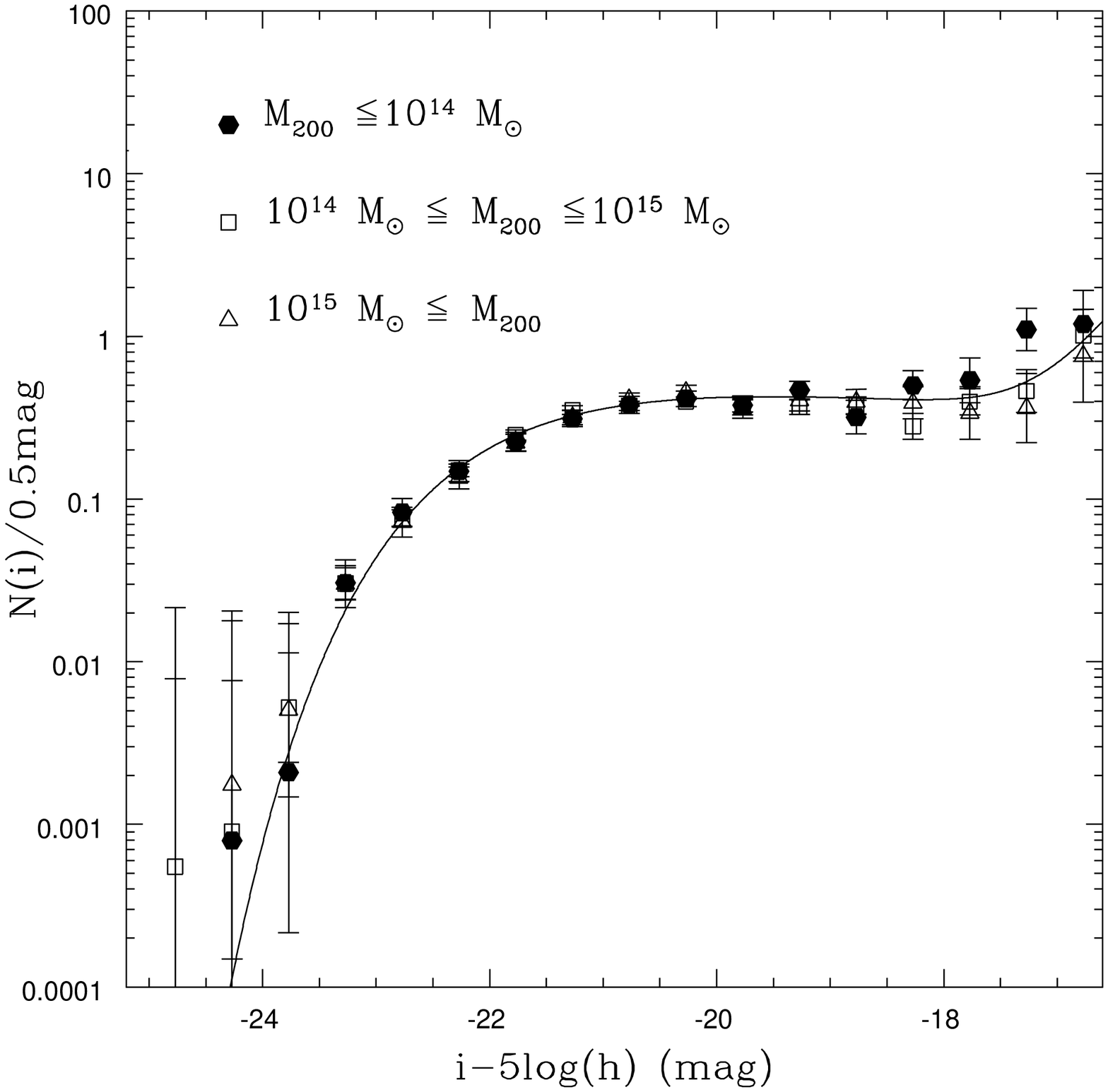}}
\end{minipage}
\end{center}
\caption{ The composite LF for the whole (left panel) and the red
(right panel) cluster galaxy populations. In both panels the filled
points are the low-mass clusters ($M_{200}/10^{14}M_{\odot} \le 1$),
the empty squares the intermediate-mass systems ($ 1 \le
M_{200}/10^{14}M_{\odot} \le 10$) and the empty triangles are the
high-mass clusters ($M_{200}/10^{14}M_{\odot} \ge 10$).  The different
mass-bin LFs are renormalized so as to ease the comparisons among
them.  The solid lines in the panels are the best fits obtained in
Popesso et al.(2006) from the X-ray selected RASS-SDSS galaxy
clusters for the corresponding whole and red cluster galaxy
populations.}
\label{lf_xf}
\end{figure*}

\section{Luminosity function shape and cluster mass}
In this section we investigate whether the lack of galaxies observed
in the high-mass systems is related to a different shape of the LFs of
clusters of different masses. The universality of the cluster LF
has been analysed in two papers of the RASS-SDSS Galaxy Cluster Survey
Series (Popesso et al. 2005a, Popesso et al. 2006). When measured
within the cluster virial radius ($r_{200}$), the shape of the LF does
not change from cluster to cluster both at the faint and at the bright
end (Popesso et al. 2006). Moreover, we observed that the cluster to
cluster variations of the LF found in the literature are due to choice
of a fixed metric apertures for all the systems. This is due to the
fact that fraction of dwarf galaxies in clusters is an increasing
function of the clustercentric distance (see also Durret et
al. 2002). To keep under control also the possible dependence between
the shape of the LF and the cluster mass, we divided our cluster
sample (with mass ranges from $10^{13}M_{\odot}$ to $4 \times
10^{15}M_{\odot}$) in three mass bins ($M_{200}/10^{14}M_{\odot} \le
1$, $ 1 < M_{200}/10^{14}M_{\odot}
\le 10$ and $M_{200}/10^{14}M_{\odot} > 10$). To increase the
statistics and study the average luminosity distribution of the
galaxies in each mass bin, we have used the SDSS photometric data to
compute a composite luminosity function (LF) by stacking the
individual cluster LFs calculated within $r_{200}$. The individual LFs
are obtained by subtracting the field number counts calculated within
an annulus around the cluster (0.2 deg with), from the number counts
in the cluster region, as described in Popesso et al. (2005a).
Following previous works, we exclude from the individual cluster LFs
the Brightest Cluster Galaxies (BCGs). The composite LF in each
mass bin is calculated by following the prescriptions of Colless
(1989, see also Popesso et al. 2005a for more details about this
method). We require at least 10 clusters contributing to each
magnitude bin of the composite LF. This requirement is fulfilled at
magnitudes brighter than the absolute magnitude limit $i-5 \log(h) \le
-16.7$ mag in all the cluster mass bins considered, while 95\% of our
clusters have magnitude limits brighter than $-18.25$ mag in the i
band. Thus, we consider galaxies down to 5.5 mag fainter than the
cluster $M^*$ in this SDSS band (Popesso et al. 2006). Moreover, we
use the stellar masses estimated by Kauffmann et al. (2003) for the
DR2 spectroscopic sample to evaluate the stellar mass range sampled
within this magnitude limit.  As shown in Fig. \ref{maz}, although the
scatter is large (0.18 dex), the magnitude cut at $-16.7$ mag
corresponds roughly to a galaxy stellar mass of $1.5 \times 10^8
M_{\odot}$. As in section~\ref{dprofs}, we distinguish between early
and late type galaxies using a SDSS color cut at $u-r = 2.22$.  In
Fig. \ref{lf_xf} we show the composite LF for the whole (left panel)
and the red (right panel) cluster galaxies populations. In both panel
the filled points are the low-mass clusters ($M_{200}/10^{14}M_{\odot}
\le 1$), the empty squares the intermediate-mass systems ($ 1 <
M_{200}/10^{14}M_{\odot} \le 10$) and the empty triangles are the
high-mass clusters ($M_{200}/10^{14}M_{\odot} > 10$). The different
mass-bin LFs are renormalized so as to ease the comparisons among
them.  The solid lines in the panels are the best fits obtained in
Popesso et al. (2006) from the X-ray selected RASS-SDSS galaxy
clusters for the corresponding whole and red cluster galaxy
populations. From Figure \ref{lf_xf} it is clear that, at
magnitudes brither than $-16.7$ mag (alternatively, for galaxy stellar
masses above $1.5 \times 10^8 M_{\odot}$), there are no significant differences
among the LFs in the different mass bins. Moreover, the best fit of
the composite LF of the X-ray selected RASS-SDSS sample provides a
very good fit to any of the considered LFs. We conclude that the
cluster LF does not depend on the cluster mass. This conclusion is
consistent with our previous findings (Popesso et al. 2006).

\begin{figure}
\begin{center}
\begin{minipage}{0.37\textwidth}
\resizebox{\hsize}{!}{\includegraphics{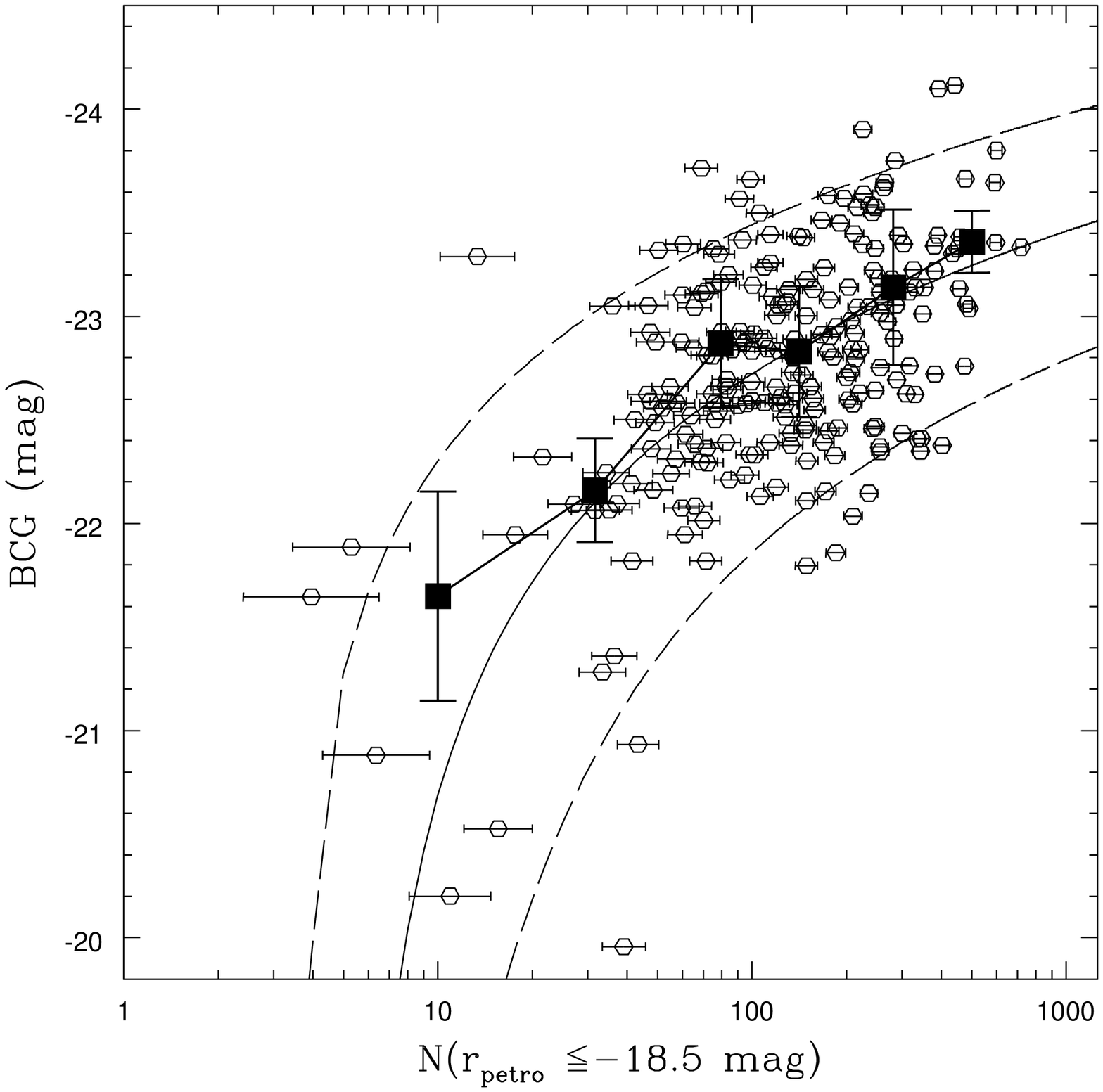}}
\end{minipage}
\begin{minipage}{0.37\textwidth}
\resizebox{\hsize}{!}{\includegraphics{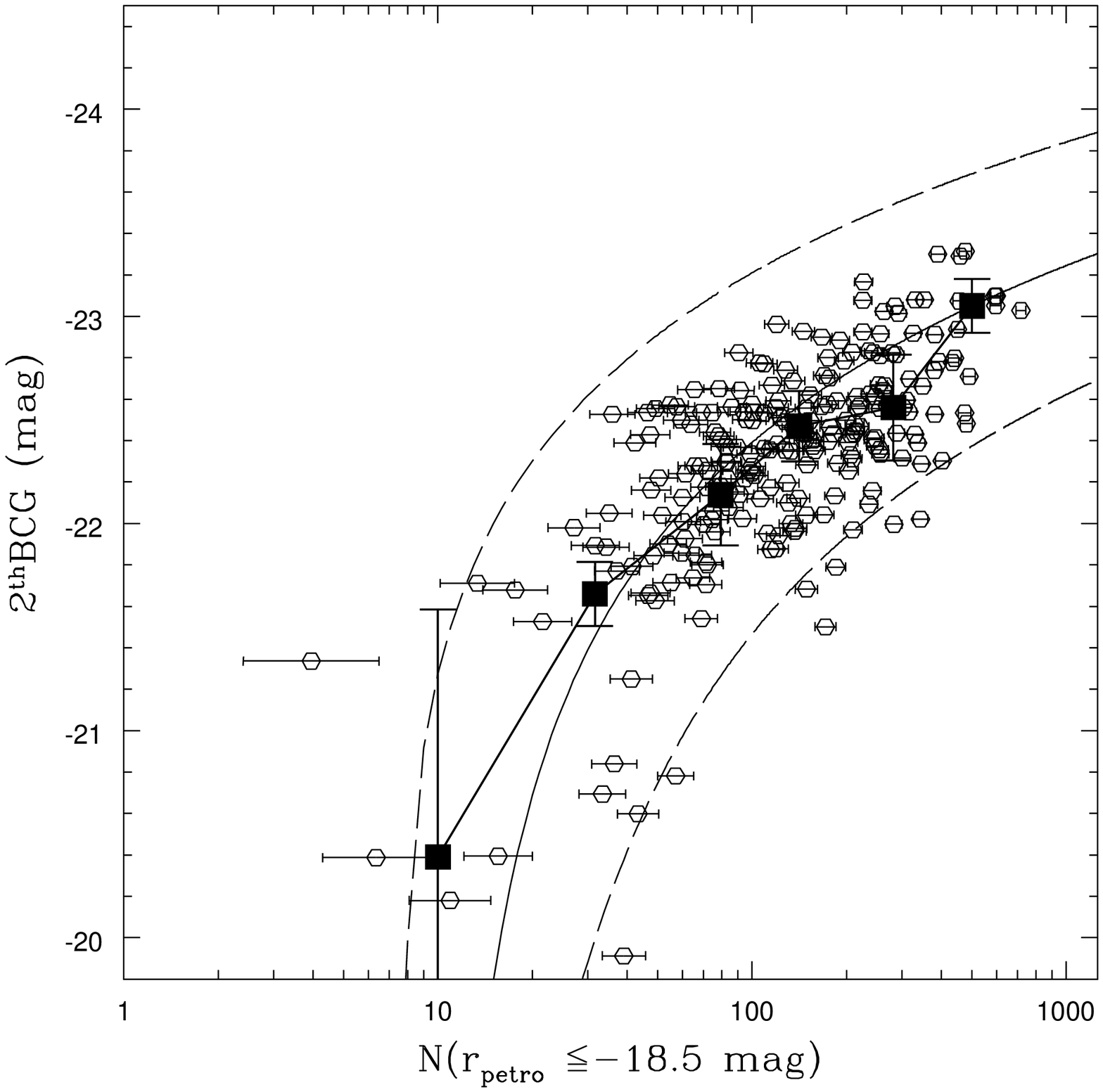}}
\end{minipage}
\begin{minipage}{0.37\textwidth}
\resizebox{\hsize}{!}{\includegraphics{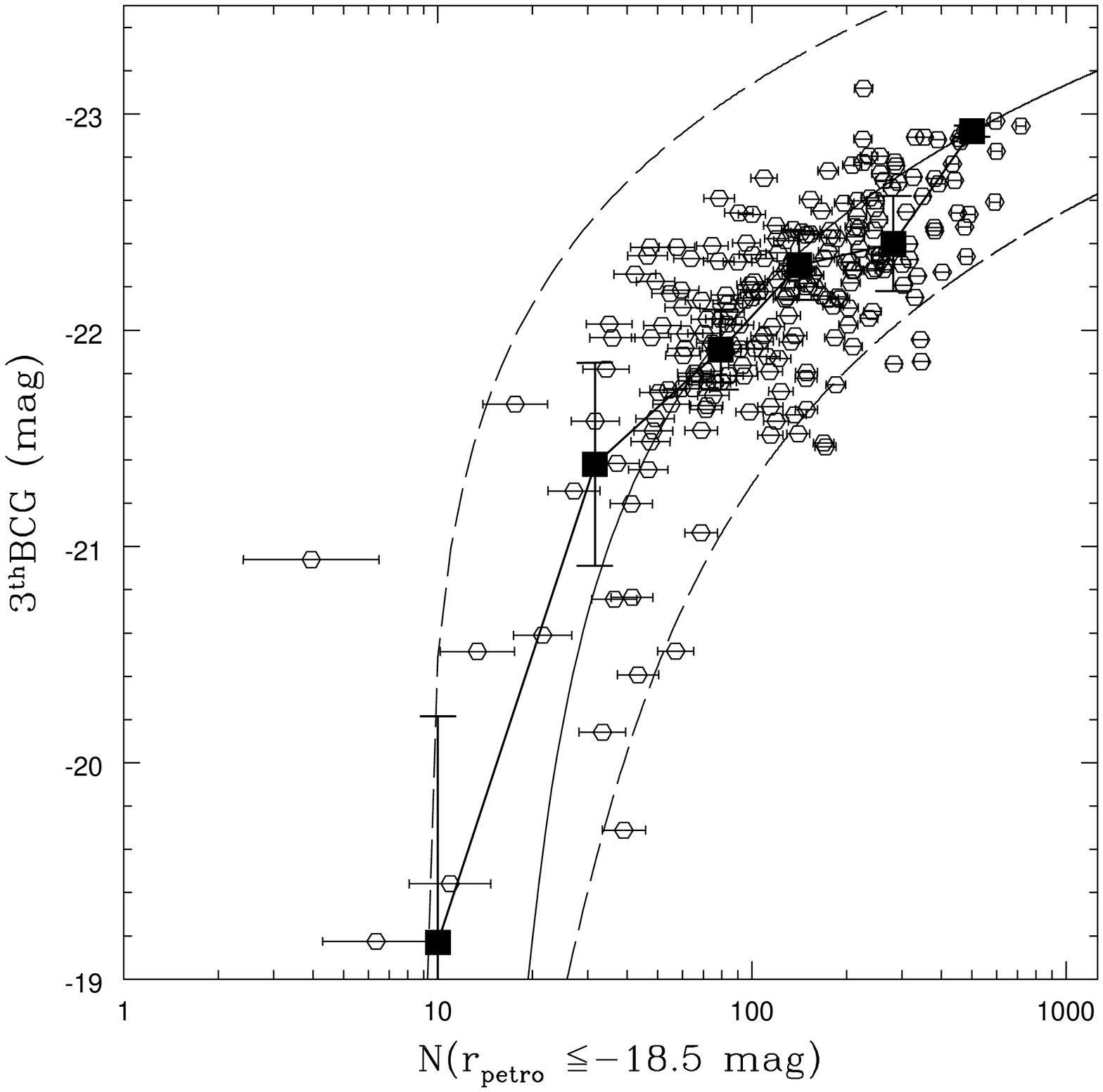}}
\end{minipage}
\end{center}
\caption{Upper panel: The magnitude of the BCGs (empty squares) within
$r_{200}$ as a function of the number of cluster galaxies within the
same radius, and with $r \le -18.5$ mag. The solid line in the plot
shows the expected BCG magnitude as estimated from the best-fitting
composite LF, as a function of the cluster normalization. The dashed
lines are the statistical uncertainties in the relation represented by
the solid line. The filled squares are the median magnitudes per bin
of $N_{gal}$.  Middle panel: same as the left panel, but for the
second brightest cluster galaxies. Bottom panel: same as the left
panel, but for the third brightest cluster galaxies. }
\label{bcg}
\end{figure}

The previous analysis is based on LFs with the BCGs excluded. Here we
examine to what extent can the BCG luminosities be considered the
high-end tail extension of the overall cluster LFs. This has been
shown {\em not} to be the case by previous investigations
(e.g. Schechter 1976, Bhavsar \& Barrow 1985).  The Schechter function
was generally found to provide a good fit to the observed galaxy
magnitude distribution as long as the very brightest galaxies, the cD
galaxies, were excluded from the fit (Schechter 1976). The exceptional
luminosities of these galaxies have often been interpreted as arising
from special processes that are not common to all galaxies, and are
particularly effective at the bottom of cluster potential wells.
Nevertheless, Lugger (1986) did not find significantly different
best-fits to the observed cluster galaxy LFs using Schechter
functions, when BCGs were or were not included in the sample.

Following Colless (1989), we normalize the cluster LFs to the number of
cluster galaxies in a common magnitude region ($r \le -18.5$ mag in
the present case, see Popesso et al. 2005a for details). Given the
number of cluster galaxies in that magnitude region and the best-fit
Schechter function of the composite LF, it is possible to calculate
the magnitude $M_r$ of the $n^{th}$ brightest cluster member as the
magnitude corresponding to $N(M_r)=n$, where $N(M_r)$ is the
analytical form of the cluster integral LF. For this we use the
best-fit obtained with a composite of two Schechter functions, after
excluding the BCGs.

The left panel of Fig.~\ref{bcg} shows the magnitude of the brightest
spectroscopically-confirmed cluster members within $r_{200}$, as a
function of the number of cluster galaxies within $r_{200}$ and with
$r \le -18.5$ mag.  The solid line shows the expected magnitude of the
brightest galaxies, as estimated from the best-fit LF, vs. the cluster
normalization. The dashed line are the statistical uncertainties in
the location of the brightest cluster member. Clearly, the estimated
magnitudes of the 1$^{st}$ ranked galaxies are consistent with the
observed values, as can be judged by considering the median of the
1$^{st}$-ranked galaxy magnitudes per $N_{gal}$ bin (filled squares in
the plot), and by the fact that 95\% of the BCGs lie within the
statistical uncertainties of the expected relation. The middle and the
right panel of Fig.~\ref{bcg} are similar to the left panel, but for
the 2$^{nd}$ and 3$^{rd}$ brightest cluster galaxies respectively.
Again, the agreement between the expected and observed magnitudes is
extremely good, and the similarity of these three plots argues
against the BCG magnitudes being an anomaly of the cluster LF.

The reason why our result disagrees with previous findings (Postman \&
Lauer 1995) must be related to the use of a double (instead of a
single) Schechter function for the fit of the observed LF, which
allows a better representation of the LF bright end. This was first
pointed out by Biviano et al. (1995) in their study of the Coma
cluster LF (see also Thompson \& Gregory 1993). The deviation of
the cluster LF from a single Schecter function was also found in the
clusters extracted from the N-body simulations combined with
semi-analytical models analysed by Diaferio et al. (1999). They
interpreted the LF shape as the effect of the large merger
cross-section of the bright and massive central galaxies.

\begin{figure}
\begin{center}
\begin{minipage}{0.5\textwidth}
\resizebox{\hsize}{!}{\includegraphics{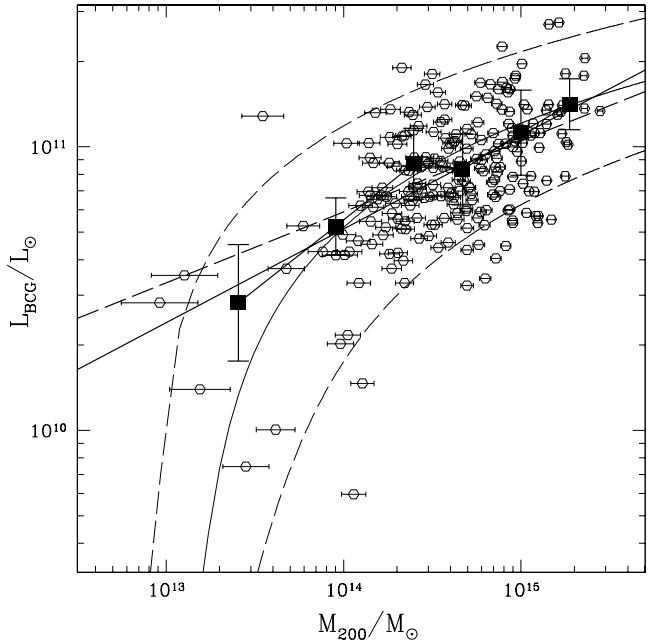}}
\end{minipage}
\end{center}
\caption{ The relation between BCG luminosity and cluster mass for our
cluster sample. Cluster masses are computed from $N_{gal}$s using our
derived HOD.  The solid line in the plot is the best-fit line obtained
with an orthogonal linear regression, $L_{BCG} \propto
M_{200}^{0.33\pm0.04}$, and the dashed line is Lin \& Mohr's (2004)
relation. Other symbols have the same meaning as in the
Fig. \ref{bcg}.}
\label{bcg_m}
\end{figure}

Our result is in agreement with the recent findings of Lin \& Mohr
(2004) and Yang et al. (2005) of a tight correlation between the BCG
luminosity and the cluster mass. In particular, in the mass range
$10^{13} \le M_{200}/M_{\odot} \le 10^{15}$, it results $L_{BCG}
\propto M_{200}^{0.25}$. The excellent agreement between Lin \& Mohr's
result and ours is demonstrated in Fig. \ref{bcg_m}. There we show the
relation between the BCG luminosity and the cluster mass of our
cluster sample, where we transformed the cluster $N_{gal}$ into
cluster masses using the HOD we derived in section~\ref{lmr}.  The
solid line in the plot is the best-fit obtained with an orthogonal
linear regression, $L_{BCG} \propto M_{200}^{0.33\pm0.04}$, and it is
in excellent agreement with the Lin \& Mohr (2004) relation (the
dashed line in the plot).

\section{The fundamental plane of cluster ellipticals}
The elliptical galaxies are the dominant population in clusters and
therefore any variation of their mass-to-light ratio as a function of
the cluster mass could contribute affecting the slope of the $N-M$ and
the $L-M$ relations. I.e. one could still have a constant ratio between
the total cluster mass and the total mass in galaxies, even for 
$\gamma < 1$ (see section~\ref{lmr}), if galaxies of given luminosity
have higher masses in higher-mass clusters. 

To investigate whether elliptical galaxies in high mass clusters have
a higher average $M/L$ than their counterparts in low mass systems, we
determine the fundamental plane (FP hereafter) of ellipticals as
traced by the spectroscopic members of each cluster within
$r_{200}$. The FP relates the effective radius of the luminosity
distribution of ellipticals, $r_e$, with their internal velocity
dispersion, $\sigma$, and their surface brightness (Djorgovsky \&
Davis 1987). If the virial radius of ellipticals is linearly
proportional to $r_e$ and their internal velocity dispersion to
the virial value, the FP effectively can be used to constrain
the mass-to-light ratio of elliptical galaxies.

For this analysis, as before, we have divided our cluster sample in
three subsamples of low-, intermediate-, and high-mass. Ellipticals
are identified within each cluster using the selection criteria of
Bernardi et al. (2003a). As a measure of the effective radius we use
the Petrosian radius $r_{50}$, which encloses 50\% of the total
Petrosian luminosity, multiplied by the square root of the ratio $b/a$
of the lengths of the minor and major axes of the observed surface
brightness profile. The SDSS spectroscopic catalog contains a measure
of the line of sight velocity dispersion which has been corrected for
aperture effects as in Bernardi et al. (2003a). In what follows, we
show the best correlation between the variables $r_e$, $\sigma$ and
$\mu=-2.5 \log[(L/2)/r_e^2]$ in the SDSS r-band. Data are fitted with
the ODRPACK routine (Akritas $\&$ Bershady 1996).

\begin{figure*}
\begin{center}
\begin{minipage}{0.49\textwidth}
\resizebox{\hsize}{!}{\includegraphics{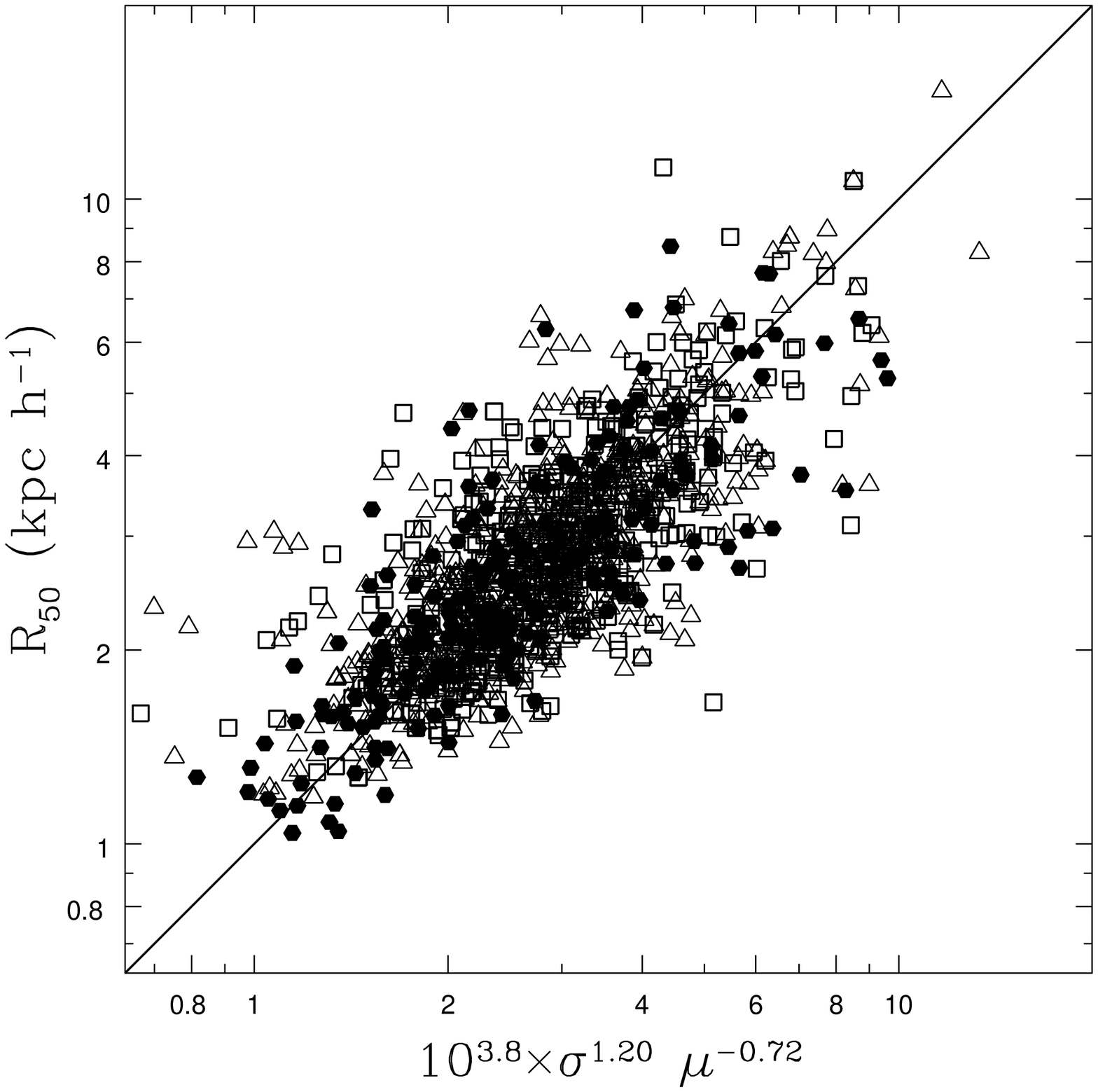}}
\end{minipage}
\begin{minipage}{0.49\textwidth}
\resizebox{\hsize}{!}{\includegraphics{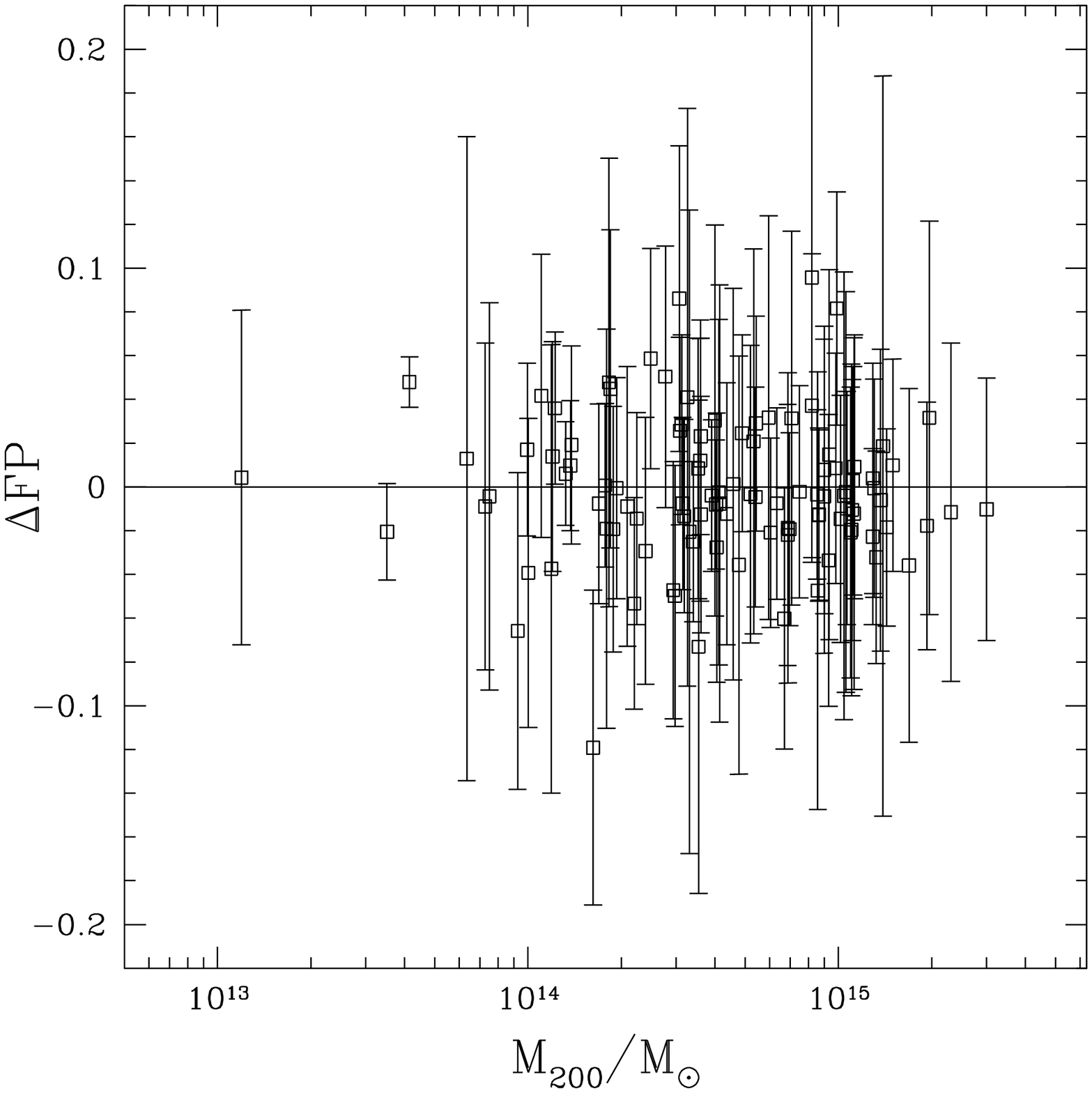}}
\end{minipage}
\end{center}
\caption{Left panel: the best-fit FP of cluster ellipticals,
relating their internal velocity dispersions
($\sigma$), effective radii ($r_e$), and surface brightnesses
within $r_e$ ($\mu$). The different symbols in the plot indicate
elliptical galaxies of clusters in different mass bins.  The
right-hand side panel shows the mean residuals from the global FP
of the elliptical populations of individual clusters as a
function of the cluster mass.}
\label{fp}
\end{figure*}

The left-hand side panel of the fig. \ref{fp} shows the best-fit FP
relating the three variables $\sigma$, $r_e$ and $\mu$; our result is
consistent with Bernardi et al. (2003b).  The different symbols in
the plot indicate elliptical galaxies of clusters in different mass
bins. We do not find any variation of the best-fit FP for
the different mass samples.  In the right panel of the same figure we
show the mean residuals from the FP of the elliptical populations of
individual clusters as a function of the cluster masses. The scatter
around the FP is $\sim 10$\% as in Bernardi et al. (2003b) and the
mean residual is consistent with zero independent of the cluster mass.

Any variation of the average mass-to-light ratio of the ellipticals
would result in a variation of their FP. The constancy of the FP
across the cluster mass range therfore implies a constant
mass-to-light ratio of the cluster ellipticals.

\section{Discussion}
Using a large sample of galaxy clusters we have shown that the number
of galaxies per unit mass is lower in clusters of higher masses,
i.e. the slope of the relation $N \propto M^{\gamma}$ is $\gamma < 1$
at the 2.5 $\sigma$ level. Our result is thus in agreement with
previous findings (see, e.g., Lin et al. 2004 and references therein)
although we find a somewhat steeper $N-M$ relation ($\gamma$ closer to
unity), because of our mass-dependent correction for projection
effects. 

>From the theoretical point of view, $\gamma < 1$ is expected.  On one
hand, hierarchical models of structure formation predict a universal
mass distributions of the subhalos (see, e.g., De Lucia et al. 2004,
and Gao et al. 2004), independent from the mass of the parent halo. As
a consequence, the number of subhalos is directly proportional to the
parent halo mass ($N \propto M$). On the other hand, including baryons
in the simulations leads to a decreasing number of galaxies per given
mass in halos of higher masses (i.e. $\gamma < 1$, see, e.g.,
Benson et al. 2000a, 200b; White et al. 2001; Berlind et al. 2003),
and of an increasing $M/L$ as a function of mass (e.g. Kauffmann
et al. 1999).  This could happen as the consequence of a decreasing
efficiency of gas cooling and star formation (see, e.g., Springel \&
Hernquist 2003; Berlind et al. 2003; Borgani et al. 2004; Kravtsov et
al. 2004), or because of an increased merger rate (White et al. 2001),
or of an increased destruction rate of galaxies (Lin et al. 2003), as
the mass of the parent halo increases.

Although we observe $\gamma < 1$ as predicted, a deeper look at other
properties of our clusters seems to be at odds with theoretical
predictions.  If mergers and/or tidal effects are responsible for
reducing the number of galaxies per given mass in clusters of higher
masses, we expect to see these processes to leave an imprint into the
cluster LFs. Instead, we find that the LF is universal, with no
dependence on the cluster mass.  Our result is at odds with Lin et
al. (2004).  The reason for this difference is unclear, but it could
be related to the different photometric bands (Lin et al.  use the
K-band), although it is difficult to see why the LFs of clusters of
different masses should look identical in four SDSS photometric bands
and different in the K-band.

Another result argues against galaxy-destruction via tidal stripping
being more efficient in higher mass clusters. Higher mass clusters are
characterized by a more concentrated number density profile (see
section~\ref{dprofs}) and a less concentrated mass density profile
(NFW; Katgert et al. 2004; Pratt \& Arnaud 2005) near the centre.  If
anything, this is consistent with a picture where galaxies are more
likely to survive near the centre of higher mass clusters, while
galaxies in lower-mass clusters are destroyed when crossing the
cluster core, because of the efficient tidal stripping resulting from
a more concentrated halo mass profile.

A lower efficiency of star formation in galaxies of higher mass
clusters would also lead to observe $\gamma < 1$ in the HOD. A
consequence of this process should be visible in a decreased $M/L$ for
the galaxies of higher mass clusters, as compared to the galaxies of
lower mass clusters. We have explored this possibility by the analysis
of the FP of cluster ellipticals.  No evidence for a variation of the
FP as a function of cluster mass was found. This result argues for a
constant $M/L$ and hence a similar star formation efficiency of
cluster ellipticals in clusters of different masses, in agreement
with the predictions of Diaferio et al. (2001), based on numerical
simulations combined with semi-analytical models of galaxy formation.
Note, however, that Springel et al. (2001) have argued that even
heavily stripped cluster galaxies obey the Faber-Jackson relation,
since the internal velocity dispersion of a stripped subhalo remains
relatively stable until it is fully disrupted. Hence the constancy of
the FP does not rule out the possibility of subhaloes stripping.

Bahcall \& Comerford (2002) have suggested that the observed
increasing $M/L$ of clusters as a function of cluster masses is a
consequence of a higher fraction of galaxies with evolved stellar
populations in higher mass clusters. There is no evidence for this in
our data (Popesso et al. 2005e). Moreover, Bahcall \& Comerford's
prediction that $M/L$ vs. $M$ would become flatter when the
photometric band is moved to longer wavelengths, is also ruled out by
our data, where we see that the relation does not change by changing
SDSS photometric band, in agreement with the results of numerical
simulations combined with semi-analytical modelling (Kauffmann et al.
1999).

How can then we reconcile the observed $N-M$ with the predictions for
a universal subhalo mass distribution? It is hard to find physical
processes capable of reducing the number of observed galaxies per
given mass, while at the same time leaving the subhalo mass
distribution, the galaxies LF, and the average galaxy mass-to-light
ratios unchanged. Hence, the most likely explanation is that the mass
distribution of the subhalos is not universal and the observed $\gamma
< 1$ for galaxies simply reflect an underlying $\gamma < 1$ for
subhalos. 

A final considerations is in order.  Our correction for projection
effects does work in the sense of changing the observed $\gamma$ of
the $N \propto M^{\gamma}$ closer to unity. The resulting $\gamma$ is
still found to be below unity, but the significance of this is not
overwhelming (2.5 $\sigma$ level). Hence it is well possible that
other insofar unapplied corrections, or, perhaps, an improved
correction for the projection effects, could make $\gamma$ consistent
with unity, thus reconciling theory and observations.

\section{Conclusions}
We have studied the $L-M$ and the $N-M$ relations in the 4 SDSS bands
g, r, i, z for a sample of 217 galaxy clusters with confirmed 3D
overdensity in the SDSS DR3 spectroscopic catalog. All the quantities
are measured within the characteristic cluster radius $r_{200}$. We
have remarked upon the direct connection between the two relations due
to the proportionality of the cluster optical luminosity and the
number of cluster galaxies.

We have studied the galaxy surface number density profile in five bins
of cluster mass and discovered that the profile has a strong
dependence on the cluster mass.  In the low and intermediate mass
systems the best fit is provided by a King profile. The core radius of
the best fit is decreasing as a function of the cluster mass, while
the central galaxy density is increasing. In the highest mass bins a
more concentrated generalized King profile or a cuspy NFW profile
provide the best fits. Using the best fit profile in each mass bin, we
have converted the observed number of cluster galaxies to the value
within the virial sphere. Since clusters of different masses exhibit
different surface density profiles, the deprojection correction
decreases with the cluster mass.  Applying this mass-dependent
correction affects the $L-M$ and $N-M$ relations, by increasing the
slope of these relations to the value of $0.92\pm0.03$. Similarly,
also the slope of the $M/L-M$ relation is affected and becomes
$0.18\pm0.04$. Hence, neglecting the dependence of the deprojection
correction on the cluster mass, leads one to underestimate the slope
of the $L_{op}-M_{200}$ and $N_{gal}-M_{200}$ relations.  Despite the
deprojection correction, the derived $N-M$ and the $L-M$ relations are
still only marginally consistent with unity, at the 2.5$\sigma$
level, i.e. direct proportionality between cluster mass and number of
cluster galaxies is not supported.

We have compared the properties of our clusters with the prediction of
the hierarchical models of structure formation. These models naturally
predict that $N \propto M^{\gamma}$ with $\gamma < 1$. This result is
generally interpreted as the indication that the galaxies in the low
mass systems are older and more luminous per unit mass than the
galaxies in high mass clusters. As a consequence, variations of the
shape of the cluster LF and of the elliptical FP with the cluster mass
are also expected. Such predicted variations are however not seen in
our data. Not only we found the LF to be the same for clusters of
different masses, but we also proved that this universal LF can be
used to accurately predict the magnitudes of the three brightest
cluster galaxies, given the LF-normalization of the clusters in which
they are located. In other words, the BCG magnitudes are consistent
with being drawn from the best-fit magnitude distribution of other
cluster galaxies. Moreover we have shown that the FP of cluster
ellipticals has the same slope in all the clusters and does not depend
on the cluster mass.

We conclude this paper with the following considerations. From the
observational point of view, the mean cluster luminosity function and
the $N-M$ or the $L-M$ relation determine completely the luminosity
distribution of cluster galaxies. The mean cluster LF constrains with
high accuracy the shape of the luminosity distribution in clusters,
while the $N-M$ relation, calculated in a given magnitude range, fixes
the normalization of the LF as a function of the cluster mass.
Forthcoming cosmological models of galaxy formation should aim at
reproducing this characteristic of the cluster galaxy populations, in
order to understand the processes of galaxy formation and evolution in
the cluster enviroment.

\vspace{2cm}

We thank the referee, Christophe Adami, for the useful comments which
helped in improving the paper. We acknowledge useful discussions with
Stefano Borgani and Simon White.  Funding for the creation and
distribution of the SDSS Archive has been provided by the Alfred
P. Sloan Foundation, the Participating Institutions, the National
Aeronautics and Space Administration, the National Science Foundation,
the U.S.  Despartment of Energy, the Japanese Monbukagakusho, and the
Max Planck Society. The SDSS Web site is http://www.sdss.org/. The
SDSS is managed by the Astrophysical Research Consortium (ARC) for the
Participating Institutions.  The Participating Institutions are The
University of Chicago, Fermilab, the Institute for Advanced Study, the
Japan Participation Group, The Johns Hopkins University, Los Alamos
National Laboratory, the Max-Planck-Institute for Astronomy (MPIA),
the Max-Planck-Institute for Astrophysics (MPA), New Mexico State
University, University of Pittsburgh, Princeton University, the United
States Naval Observatory, and the University of Washington.

\end{document}